\documentclass[english]{article}
\usepackage[T1]{fontenc}
\usepackage[latin9]{inputenc}
\usepackage{geometry}
\usepackage{babel}
\usepackage{url}
\usepackage{amsmath}
\usepackage{amssymb}
\usepackage{graphicx}
\usepackage{esint}
\usepackage[unicode=true,pdfusetitle,
 bookmarks=true,bookmarksnumbered=false,bookmarksopen=false,
 breaklinks=false,pdfborder={0 0 1},backref=false,colorlinks=false]
 {hyperref}

\makeatletter
\numberwithin{equation}{section}
\newcommand{\lyxaddress}[1]{
\par {\raggedright #1
\vspace{1.4em}
\noindent\par}
}

\@ifundefined{showcaptionsetup}{}{%
 \PassOptionsToPackage{caption=false}{subfig}}
\usepackage{subfig}
\makeatother

\begin{document}

\title{Simulating collisions of thick nuclei \\
in the color glass condensate framework}

\author{Daniil Gelfand, Andreas Ipp, David M\"{u}ller}

\maketitle

\lyxaddress{\begin{center}
\emph{Institut f\"{u}r Theoretische Physik, Technische Universit\"{a}t
Wien, }\\
\emph{Wiedner Hauptstr. 8-10, A-1040 Vienna, Austria}\\
\emph{E-Mail}: \href{mailto:gelfand@hep.itp.tuwien.ac.at}{gelfand@hep.itp.tuwien.ac.at},
\href{mailto:ipp@hep.itp.tuwien.ac.at}{ipp@hep.itp.tuwien.ac.at},
\href{mailto:david.mueller@tuwien.ac.at}{david.mueller@tuwien.ac.at}
\par\end{center}}
\begin{abstract}
We present our work on the simulation of the early stages of heavy-ion
collisions with finite longitudinal thickness in the laboratory frame
in 3+1 dimensions. In particular we study the effects of nuclear thickness
on the production of a glasma state in the McLerran-Venugopalan model
within the color glass condensate framework. A finite thickness enables
us to describe nuclei at lower energies, but forces us to abandon
boost invariance. As a consequence, random classical color sources
within the nuclei have to be included in the simulation, which is
achieved by using the colored particle-in-cell method. We show
that the description in the laboratory frame agrees with boost-invariant
approaches as a limiting case. Furthermore we investigate collisions
beyond boost invariance, in particular the pressure anisotropy in
the glasma.

\tableofcontents{}

\newpage{}
\end{abstract}

\section{Introduction}

Heavy ion collisions allow us to study strongly interacting matter
in a deconfined phase, the quark gluon plasma. In search for a critical
point in the QCD phase diagram, experiments cover a wide range of
collision energies, from very high energies at RHIC and LHC, down
to lower energies in the Beam Energy Scan program of RHIC \cite{Adamczyk:2013gw}
and at future programs at GSI FAIR and JINR NICA. The early times
of heavy ion collisions  can be appropriately described  in the
color glass condensate (CGC) framework \cite{Iancu:2002xk,Iancu:2001yq}. 

The CGC framework models ultrarelativistic, highly Lorentz contracted
nuclei in terms of an effective classical field theory. Hard partons
are described as  color charges, which act as sources for the soft
partons in terms of classical non-Abelian gauge fields due to gluon
saturation. The distribution of the color charges of very large nuclei
is given by the McLerran-Venugopalan (MV) model \cite{MV1,MV2}. More
recent sophisticated models such as IP-Glasma base the color charge
distribution on fits to deep-inelastic scattering data \cite{Schenke:2012fw,Schenke:2012wb}.
As a result of the collision the glasma is produced \cite{Gelis:2012ri},
which can be studied by numerically solving the Yang-Mills equations.

A common simplification is to assume infinitely thin incoming nuclei,
which leads to a single collision point in time and consequently to
boost invariance. This reduces the system to effectively 2+1 dimensions
\cite{Krasnitz:1998ns,Lappi:2003bi,Kovner:1995ja}.  In this formulation
the gauge fields are rapidity independent by assumption. It is possible
to introduce rapidity dependence by including boost invariance breaking
fluctuations on top of boost-invariant background fields \cite{Gelis:2013rba,Fukushima:2011nq,Berges:2012cj}.
However, the initial conditions and evolution of the background fields
are still formulated in a boost-invariant way. Simulations of the
early stages of heavy-ion collisions using the CGC framework and real-time
lattice gauge theory have been highly successful in describing particle
multiplicities \cite{Schenke:2013dpa} and the azimuthal anisotropy
\cite{Schenke:2015aqa,Dusling:2014oha}. Studies of somewhat later
time intervals involving isotropization and thermalization of the
glasma have been undertaken using classical-statistical lattice gauge
theory with \cite{Gelfand:2016prm} and without fermions \cite{Berges:2008zt,Berges:2013fga,Romatschke:2006nk,Berges:2014yta},
hard loop approximation \cite{Ipp:2010uy,Romatschke:2006wg,Rebhan:2008uj,Rebhan:2009ku,Attems:2012js}
as well as kinetic theory \cite{Blaizot:2011xf,York:2014wja,Kurkela:2014tea,Kurkela:2015qoa}.
Within the CGC framework, there has also been progress in finding
analytical solutions for the gauge fields in the forward light cone
using expansions in small $\tau$ \cite{Chen:2015wia,Li:2016eqr,Fries:2006pv}.
Because of the infinitesimal thickness of nuclei in all the boost-invariant
approaches, the evolution of the color sources can be solved analytically.

Nontrivial evolution of color sources in the form of charged particles
can be simulated using the colored particle-in-cell method (CPIC)
method. It combines classical field dynamics described by real-time
non-Abelian lattice gauge theory \cite{PhysRevD.11.395} with classical
colored particle dynamics based on Wong's equations \cite{Wong}.
It is a non-Abelian generalization of the particle-in-cell (PIC) method
for the simulation of Abelian plasmas \cite{Verboncoeur2005}. CPIC
has been successfully applied to hard-thermal-loop simulations \cite{Hu:1996sf,Moore:1997sn},
the investigation of plasma instabilities \cite{Strickland:2007,Dumitru:2005hj}
and to jet energy loss \cite{Schenke2009} in the quark-gluon plasma.
Apart from pioneering work \cite{Poschl:1998px,Poschl:1999dm,Poschl:2000ip}
for very small transversal lattices this approach has not been used
yet to investigate the collision itself.

In this paper, we simulate the collision of two nuclei with finite
thickness in the laboratory frame in 3+1 dimensions in the CGC framework.
A finite nuclear thickness enables us to describe nuclei at lower
energies. Without a well-defined collision point, we have to drop
the assumption of boost invariance for the fields in the forward light cone.
As a consequence of an extended collision, we cannot describe the
evolution of the color sources analytically and we are forced to
include the color charges as dynamical degrees of freedom in the
simulation as they traverse the evolving overlap region of the two
nuclei. In studying lower collision energies, we are probing the limits
of applicability of the CGC framework, which becomes an accurate effective
description of QCD only at infinitely high energies. The goal of
this work is to show that a description in the laboratory frame using
CPIC is viable and can reproduce well-known results of boost-invariant
classical Yang-Mills simulations in the limit of small longitudinal
thickness of the nuclei. For simplicity we restrict ourselves to collisions
in the MV model, which describes ultrarelativistic nuclear matter
infinitely extended in the transversal directions, and to the gauge
group SU(2). We do not take into account other possible effects that
might come into play if the CGC picture is applied to lower energies,
but simply approach this region as a first step by varying the thickness
of the incoming nuclei.

This paper is organized as follows: In Sec.~\ref{sec:Colored-particle-in-cell}
we describe the CPIC method for heavy-ion collisions. We discuss the
equations of motion for the fields and color charges and initial conditions
in the laboratory frame. In Sec.~\ref{sec:Numerical-results} we
present our numerical results for collisions in the MV model with
finite thickness. We investigate the structure of the fields in the
forward light cone created during the collision and compare them to
the initial conditions used in boost-invariant simulations. We recover
the usual result of pressure anisotropy and investigate the energy
conservation in the system.\newpage{}

\section{Colored particle-in-cell method for heavy-ion collisions\label{sec:Colored-particle-in-cell}}

\begin{figure}
\begin{centering}
\includegraphics[scale=0.9]{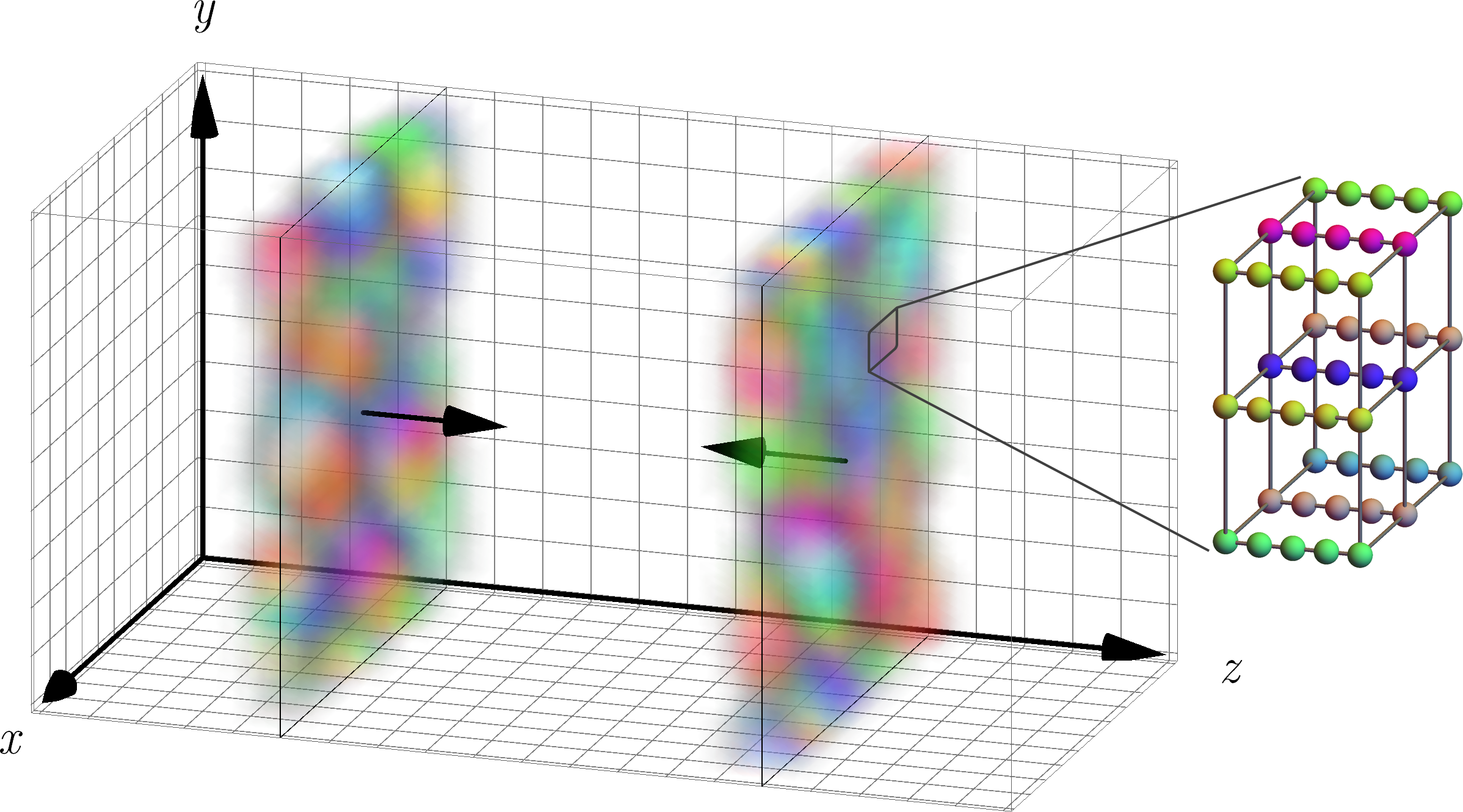}
\par\end{centering}

\protect\caption{Schematic overview of a heavy-ion collision modeled with CPIC in the
laboratory frame. The random color charge densities of the nuclei
are modeled by placing color charge carrying particles (here depicted
as small spheres) into each cell along the longitudinal direction.
These particles move continuously in the longitudinal direction, but
are fixed to the grid points in the transversal plane.\label{fig:color-sheets}}
\end{figure}

\global\long\def\Np{N_{\mathrm{p}}}

The model of a heavy-ion collision which we implement in this work
is that of two sheets of color fields and charges, each occupying
a two-dimensional plane, colliding with each other at the speed of
light in the laboratory frame. The sheets modeling Lorentz contracted
nuclei as depicted in Fig.~\ref{fig:color-sheets} have a finite
extent in the longitudinal direction in which they propagate and a
largely random transversal color structure. On the other hand, in
our setup the longitudinal color structure is assumed to be coherent,
spreading a given color configuration over the complete thickness
of a nucleus. Each of the sheets consists of two contributions, a
charge distribution and its corresponding classical fields. The charge
distributions are chosen according to the McLerran-Venugopalan model
and are not directly participating in the collision dynamics while
being tied to the light cone. In our CPIC approach we model these
charges as classical particles with a non-Abelian color charge. Following
the core assumptions of the CGC framework, the charges generate classical
gluon fields, which travel alongside them in the sheet and are responsible
for the creation of matter during and after the collision. The dynamics
of these fields are consistently described by Yang-Mills equations
without any approximations.

The CPIC method simulates the evolution of colored point charges in
continuous phase space coupled to non-Abelian gauge fields on a discrete
lattice. In each simulation step, the equations of motion for particles
and fields are solved alternately. Currents for the field equations
are obtained by interpolating the motion of charges to the lattice,
while interpolating the discretized fields back to the continuous
particle positions gives rise to forces and parallel transport. The
simulation volume is modeled as a three-dimensional grid with $N_{L}\cdot N_{T}^{2}$
cells, where $N_{L}$ and $N_{T}$ are the number of cells in the
longitudinal and transversal directions respectively with spatial
lattice spacing $a_{s}$ and time step $a_{t}$. In each cell we define
the electric fields $E_{x,i}$, the gauge links $U_{x,i}$, and the
charge and current densities $\rho_{x}$ and $j_{x,i}$, where $x$
denotes the lattice site of the cell and $i\in\{1,2,3\}$ is a vector
index. The box is periodic in the transversal directions and fixed
boundary conditions are used for the longitudinal direction.

The initial conditions for the fields are generated from the charge
densities $\rho_{1}$ and $\rho_{2}$ of the two nuclei. This step
is described in Sec.~\ref{sec:Initial-conditions-fields}. The exact
form of $\rho_{(1,2)}$ depends on the model used to describe the
nuclei. We choose the longitudinal separation of the nuclei such that
the fields do not overlap in the beginning. The color charge densities
$\rho_{(1,2)}$ are then used to sample the particle charges. We place
$\Np$ particles in each cell and apply the charge refinement algorithm
from Sec.~\ref{sub:Charge-refinement} to get a smooth distribution
of color charges among the particles. The fields and particles are
then evolved via the lattice equations of motion (see Sec.~\ref{sub:Field-equations-of})
and the nearest-grid-point interpolation method (see Sec.~\ref{sub:Nearest-grid-point-interpolation}).
Consequently, the Gauss constraint is fulfilled throughout the simulation
of the system. Similar to most collision simulations in the color
glass condensate framework we do not take the backreaction from the
fields onto the particles into account (apart from parallel transport
of the charges), i.e.~the particles' velocity is held constant at
the speed of light. As a consequence the particles act as a reservoir
of energy for the fields and total energy is not conserved. The maximum
simulation time is limited by the longitudinal length of the simulation
box, since the nuclei are continuously moving in the longitudinal
direction and will reach the end of the box after some time. The
color charge density $\rho^{a}(x)$ is treated as a random variable
following a probability functional $W[\rho]$ given by the MV model.
 Observables are recorded during the simulation and then averaged
using a number of different initial charge densities $\rho_{(1,2)}$
according to $W[\rho]$.

\subsection{Field equations of motion\label{sub:Field-equations-of}}

\global\long\def\t{\dagger}
\global\long\def\p{\partial}
In this section we review the standard lattice Yang-Mills equations
of motion. We start by discretizing the continuum Yang-Mills action
\begin{equation}
S=\intop d^{4}x\left[-\frac{1}{2}\mbox{tr}(F_{\mu\nu}F^{\mu\nu})+2\mbox{tr}(j_{\mu}A^{\mu})\right],
\end{equation}
with current density $j_{\mu}$, gauge field $A_{x,\mu}\equiv A_{\mu}(x)=A_{x,\mu}^{a}t^{a}$,
and field strength tensor $F_{\mu\nu}=\p_{\mu}A_{\nu}-\p_{\nu}A_{\mu}-ig\left[A_{\mu},A_{\nu}\right],$
on a hypercubic lattice taking advantage of the lattice gauge formalism
in Minkowski space. The gauge links $U_{x,i}$ and $U_{x,0}$ at
the lattice site $x$ are defined by
\begin{eqnarray}
U_{x,i} & = & \exp(iga_{s}A_{x,i}),\\
U_{x,0} & = & \exp(iga_{t}A_{x,0}),
\end{eqnarray}
with the temporal and spatial lattice spacings $a_{t}$ and $a_{s}$.
We also define $U_{x,-\mu}\equiv U_{x-\mu,\mu}^{\t}$ and the plaquette
variables
\begin{eqnarray}
U_{x,ij} & = & U_{x,i}U_{x+i,j}U_{x+j,i}^{\t}U_{x,j}^{\t}\simeq\exp\left(iga_{s}^{2}F_{x,ij}\right),\\
U_{x,0i} & = & U_{x,0}U_{x+0,i}U_{x+i,0}^{\t}U_{x,i}^{\t}\simeq\exp\left(iga_{s}a_{t}F_{x,0i}\right),
\end{eqnarray}
where $F_{x,ij}$ and $F_{x,0i}$ are components of the non-Abelian
field-strength tensor. The continuum action can then be approximated
as
\begin{equation}
S\simeq S_{YM}+S_{J},
\end{equation}
with the Yang-Mills part
\begin{equation}
S_{YM}=\frac{a_{s}}{g^{2}a_{t}}\sum_{x}\left(\sum_{i=1}^{3}\mbox{tr}\left[U_{x,0i}+U_{x,0i}^{\t}\right]-\frac{1}{2}\left(\frac{a_{t}}{a_{s}}\right)^{2}\sum_{i=1}^{3}\sum_{j=1}^{3}\mbox{tr}\left[U_{x,ij}+U_{x,ij}^{\t}\right]\right)+C,
\end{equation}
and the source terms
\begin{equation}
S_{J}=2a_{s}^{3}a_{t}\sum_{x}\left(\mbox{tr}\left[\rho_{x}A_{x,0}\right]-\sum_{i=1}^{3}\mbox{tr}\left[j_{x,i}A_{x,i}\right]\right),
\end{equation}
where $C$ is an irrelevant constant. By varying the discretized action
with respect to the gauge fields $A_{x,\mu}$ and employing the temporal
gauge $U_{x,0}=\mathbf{1}$, which corresponds to $A_{0}=0$, we obtain
the discretized equations of motion. For our numerical approach we
choose the electric field $E_{x,i}\equiv F^{i0}$ and the spatial
gauge links $U_{x,i}$ as our degrees of freedom. The equations can
be solved numerically using a leap-frog scheme, where the electric
fields $E_{x,i}$ and charge density $\rho_{x}$ are evaluated at
whole time steps $t_{n}=na_{t}$, while the gauge links $U_{x,i}$
and current density $j_{x,i}$ are evaluated at half time steps $t_{n+\frac{1}{2}}=(n+\frac{1}{2})a_{t}$.
The discretized equations then read

\begin{equation}
U_{x,i}(t+\frac{a_{t}}{2})=\exp\left(-ia_{t}ga_{s}E_{x,i}(t)\right)U_{x,i}(t-\frac{a_{t}}{2}),\label{eq:eom_4b}
\end{equation}

\begin{equation}
E_{x,i}^{a}(t+a_{t})=E_{x,i}^{a}(t)+\frac{a_{t}}{ga_{s}^{3}}\sum_{j\neq i}2\mbox{Im}\,\mbox{tr}\left[t^{a}U_{x,ij}(t+\frac{a_{t}}{2})+t^{a}U_{x,i-j}(t+\frac{a_{t}}{2})\right]-a_{t}j_{x,i}^{a}(t+\frac{a_{t}}{2}).\label{eq:eom_4a}
\end{equation}
Since the Gauss constraint
\begin{equation}
\sum_{i=1}^{3}\frac{E_{x,i}(t)-U_{x-i,i}^{\t}(t-\frac{a_{t}}{2})E_{x-i,i}(t)U_{x-i,i}(t-\frac{a_{t}}{2})}{a_{s}}=\rho_{x}(t)\label{eq:gauss_4}
\end{equation}
must be preserved at every time step, the charge density $\rho_{x}$
and the current density $j_{x,i}$ must obey the covariant continuity
equation, i.e.
\begin{equation}
\frac{\rho_{x}(t)-\rho_{x}(t-a_{t})}{a_{t}}+\sum_{i=1}^{3}\frac{j_{x,i}(t-\frac{a_{t}}{2})-U_{x-i,i}^{\t}(t-\frac{a_{t}}{2})j_{x-i,i}(t-\frac{a_{t}}{2})U_{x-i,i}(t-\frac{a_{t}}{2})}{a_{s}}=0.\label{eq:continuity_1}
\end{equation}

\subsection{Particle equations of motion and interpolation\label{sub:Nearest-grid-point-interpolation}}

In the CGC framework hard partons are described by classical color
sources in terms of the charge density $\rho_{x}$. Within our simulations
we sample $\rho_{x}$ by a number of pointlike particles carrying
color charge. The interpolation step reconstructs the charge density
from the particle charges and continuous positions. In the transverse
plane of the heavy ion, we place one particle per cell in order to
match the resolution of the grid. As we will see later, multiple particles
per cell in the propagating direction are needed for better resolution
of the longitudinal profile. While the colored particles move through
the grid, they induce color currents $j_{x}^{a}$, which are used
to evolve the gauge fields via the lattice equations of motion (\ref{eq:eom_4b})
and (\ref{eq:eom_4a}). A main requirement is that the Gauss constraint
(\ref{eq:gauss_4}) must be satisfied at all times. This can be accomplished
by making sure that the currents generated by the particle movement
satisfy the covariant continuity equation (\ref{eq:continuity_1}).
This is the main idea behind charge-conserving methods, which are
commonly used in Abelian PIC simulations \cite{Esirkepov2001}. 

One of the assumptions of the color glass condensate framework is
that the nuclei involved in the collision can be thought of as recoilless
sources moving at the speed of light. The charges of the nuclei pass
through each other without loss of energy or change of momentum, i.e.~the
particle trajectories are fixed. The longitudinal particle positions
$z(t)$ are simply updated with
\begin{equation}
z(t+a_{t})=z(t)+va_{t},\label{eq:position_update}
\end{equation}
with the velocity $v=\pm1$. This renders the interpolation problem
one dimensional in the longitudinal direction. Using this simplification
the continuity equation (\ref{eq:continuity_1}) reads
\begin{equation}
\frac{\rho_{x}(t)-\rho_{x}(t-a_{t})}{a_{t}}+\frac{j_{x}(t-\frac{a_{t}}{2})-U_{x-e_{z}}^{\t}(t-\frac{a_{t}}{2})j_{x-e_{z}}(t-\frac{a_{t}}{2})U_{x-e_{x}}(t-\frac{a_{t}}{2})}{a_{s}}=0,\label{eq:cont_2}
\end{equation}
where we choose $i=z$ as the longitudinal direction and drop the
direction indices.

In the simulation, we also need to interpolate the continuous particle
positions to the fixed lattice points of the charge density $\rho_{x}(t)$.
In this work, we implement a simple interpolation method  called
the nearest-grid-point (NGP) method \cite{Moore:1997sn}. In the NGP
method a particle charge $Q(t)$ at position $x(t)$ is  fully mapped
to the closest lattice point $n$. The charge density contribution
at this point from one particle is then given by
\begin{equation}
\rho_{n}(t)=\frac{Q(t)}{a_{s}^{3}}.\label{eq:ngp_charge_interpolation}
\end{equation}
As the charge moves through the grid, the charge density only
changes when the particle crosses the boundary in the middle of a
cell such that its nearest-grid-point changes. These boundaries can
be formally defined as the ones separating two cells on a lattice,
which is shifted by half a lattice spacing (Wigner-Seitz lattice),
with lattice points now marking the center and not the edges of each
cell. A current is only induced at such a boundary crossing. Evaluating
the one dimensional continuity equation (\ref{eq:cont_2}) at $x=n$
and at $x=n+1$ and requiring that the only nonzero current is $j_{n}(t-\frac{a_{t}}{2})$,
we find for a right-moving particle that moves from position $n$
to $n+1$ from time $t-a_{t}$ to $t$ the following current and updated
charge:
\begin{equation}
j_{n}(t-\frac{a_{t}}{2})=\frac{a_{s}}{a_{t}}\frac{Q(t-a_{t})}{a_{s}^{3}},\label{eq:right_move_current}
\end{equation}
\begin{equation}
Q(t)=U_{n}^{\t}(t-\frac{a_{t}}{2})Q(t-a_{t})U_{n}(t-\frac{a_{t}}{2}).\label{eq:right_move_transport}
\end{equation}
For the case of a left-moving particle from position $n$ to $n-1$
we get
\begin{equation}
j_{n-1}(t-\frac{a_{t}}{2})=-\frac{a_{s}}{a_{t}}\frac{Q(t-a_{t})}{a_{s}^{3}},\label{eq:left_move_current}
\end{equation}
\begin{equation}
Q(t)=U_{n-1}(t-\frac{a_{t}}{2})Q(t-a_{t})U_{n-1}^{\t}(t-\frac{a_{t}}{2}).\label{eq:left_move_transport}
\end{equation}
Equations (\ref{eq:right_move_transport}) and (\ref{eq:left_move_transport})
take care of the parallel transport of the charges.

The current generated by the NGP scheme can give rise to a lot of
numerical noise due to peaklike currents being induced only at certain
time steps. However, we can circumvent this problem by initializing
multiple particles per cell and by employing charge refinement procedures
(see Sec.~\ref{sub:Charge-refinement}). These improvements allow
us to simulate sufficiently accurate currents on the grid. Another
way to address this issue is to use more sophisticated interpolation
schemes such as the cloud-in-cell (CIC) interpolation, which is standard
for Abelian PIC simulations and also has been developed for the CPIC
method \cite{Strickland:2007}.

\subsection{Initial conditions\label{sec:Initial-conditions-fields}}

\begin{figure}
\begin{centering}
\includegraphics[scale=0.18]{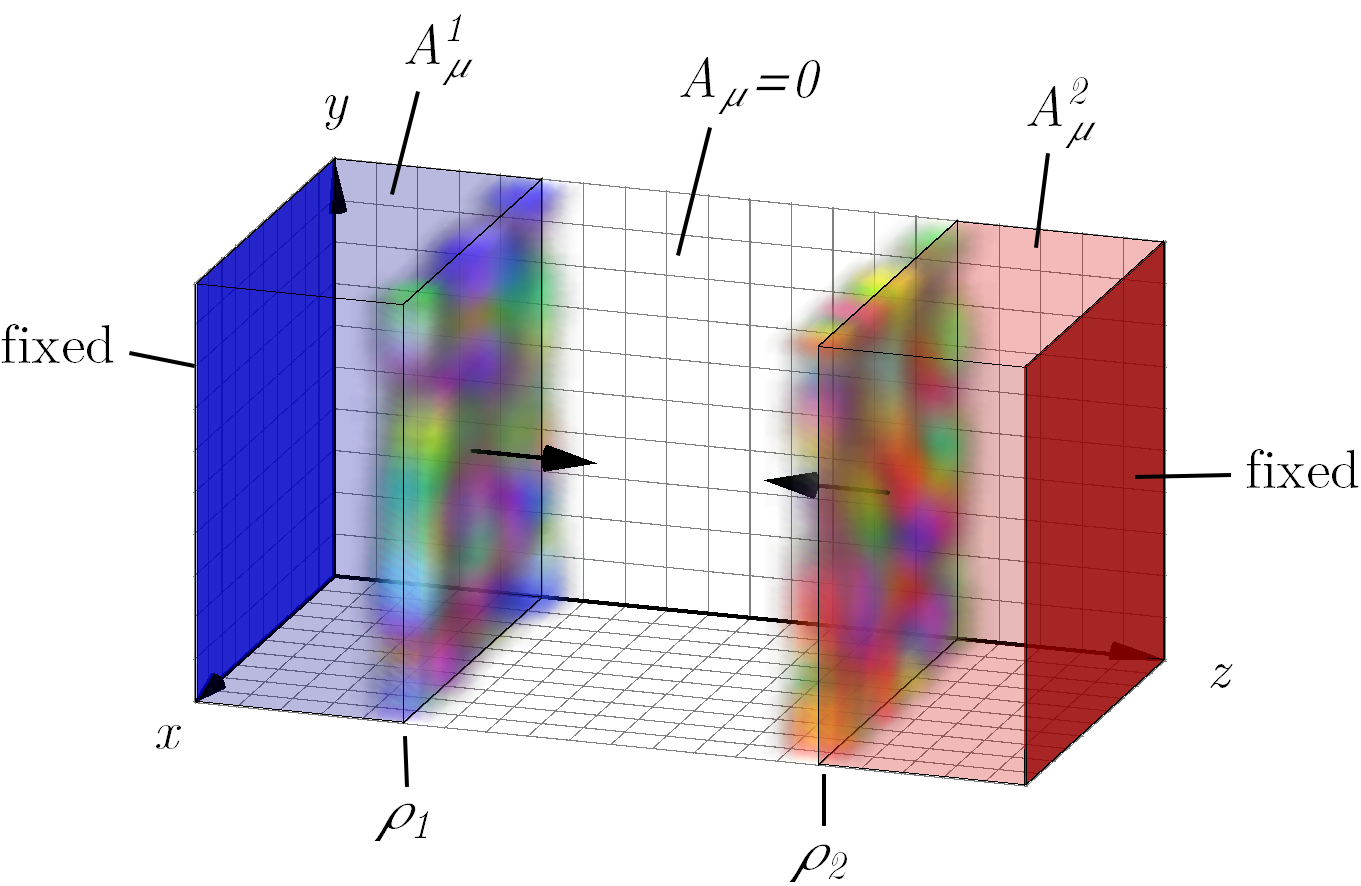}
\par\end{centering}

\protect\caption{Schematic overview of the initial conditions in temporal gauge before
the collision. The color charge densities $\rho_{(1,2)}$ of the colliding
nuclei are depicted as colorful clouds. The gauge field $A_{\mu}$
in the center region of the box is exponentially close to zero. Behind
each nucleus the fields asymptotically approach the pure gauge configurations
$A_{\mu}^{\mathrm{(1,2)}}$ depicted as blue and red transparent regions.
At the longitudinal boundaries of the simulation box the gauge fields
are fixed to the static pure gauge configurations. \label{fig:temporal-gauge-overview}}
\end{figure}

As a model of a single nucleus, we want to construct a propagating
solution with given color charge $\hat{\rho}^{a}(x_{T})$ (not to
be confused with $\rho_{x}$, the charge density in three dimensions)
in the transverse plane as given by the MV model with $x_{T}=(x,y)$
denoting the transverse coordinates. The boost-invariant case assumes
that the nucleus is infinitely Lorentz contracted in the longitudinal
direction and therefore described by a color current,
\begin{equation}
J^{a\mu}=\delta(z-t)\hat{\rho}^{a}(x_{T})s^{\mu},\label{eq:initial_conditions_current_bi}
\end{equation}
with $s^{\mu}\equiv\left(1,0,0,1\right)^{\mu}$ for a random transverse
color charge configuration $\hat{\rho}^{a}(x_{T})$ that travels at
the speed of light in the positive $z$-direction. The restriction
we release is the requirement of an infinitely thin nucleus, by spreading
the color charge along the longitudinal direction. It is possible
to find a corresponding consistent field configuration such that charge
and fields both propagate together at the speed of light. 

It is easiest to set up the solution in Lorenz gauge $\partial_{\mu}A^{a\mu}=0$.
We use the following ansatz for the four-current $J^{a\mu}=(\rho^{a},j_{i}^{a})$
and vector potential $A^{a\mu}$:
\begin{eqnarray}
J^{a\mu} & = & f(z-t)\hat{\rho}^{a}(x_{T})s^{\mu},\label{eq:Jansatznonabelian}\\
A^{a\mu} & = & f(z-t)\hat{\varphi}^{a}(x_{T})s^{\mu}.\label{eq:Aansatznonabelian}
\end{eqnarray}
The envelope $f(z-t)$ is arbitrary, and we will choose a Gaussian
profile
\begin{equation}
f(z-t)=\frac{1}{\sqrt{2\pi}\sigma}e^{-\frac{(z-t)^{2}}{2\sigma^{2}}},
\end{equation}
with a given width $\sigma$, which is proportional to the thickness
of the nucleus in the laboratory frame.\footnote{In the boost-invariant case it is common to write the expressions
$J^{a\mu}$ and $A^{a\mu}$ a bit differently using light cone coordinates
$x^{\pm}=\frac{1}{\sqrt{2}}(t\pm z)$. One would then write Eq.~(\ref{eq:initial_conditions_current_bi})
as $J^{a\mu}=\delta(x^{-})\hat{\rho}^{a}(x_{T})\bar{s}^{\mu}$, where
$\bar{s}^{\mu}=\frac{1}{\sqrt{2}}\left(1,0,0,1\right)^{\mu}$ is the
unit vector in the $+$ direction. Applying this to nuclei with finite
thickness we would have $J^{a\mu}=\bar{f}(x^{-})\hat{\rho}^{a}(x_{T})\bar{s}^{\mu}$
and $A^{a\mu}=\bar{f}(x^{-})\hat{\varphi}^{a}(x_{T})\bar{s}^{\mu}$,
where $\bar{f}(x^{-})$ is a Gaussian profile with thickness parameter
$\bar{\sigma}$. Our convention differs from this by introducing the
thickness parameter $\sigma$ in the laboratory frame coordinates
instead of the light cone coordinate frame. The different conventions
for the widths $\sigma$ and $\bar{\sigma}$ are geometrically related
by $\sqrt{2}\bar{\sigma}=\sigma$. In the end it does not matter which
convention one uses, but one should be aware that a finite width in
the light cone frame differs from the width in the laboratory frame
by a factor of $\sqrt{2}$.} Plugging the ansatz into the non-Abelian Maxwell equations 
\begin{equation}
D_{\mu}^{ab}F^{b\mu\nu}=J^{a\nu},
\end{equation}
 nonlinear terms vanish due to $s^{\mu}s_{\mu}=0$ and the time dependence
drops out because of $s^{\mu}\partial_{\mu}f(z-t)=0$. As a consequence
one is left with the Poisson equation
\begin{equation}
-\Delta_{T}A^{a\nu}\equiv-\left(\frac{\partial^{2}}{\partial x^{2}}+\frac{\partial^{2}}{\partial y^{2}}\right)A^{a\nu}=J^{a\nu},\label{eq:Poisson_eq}
\end{equation}
which is solved in Fourier space for each color component separately
and formally denoted using the inverse Laplace operator $\Delta_{T}^{-1}=\left(\nabla_{T}^{2}\right)^{-1}$
by 
\begin{equation}
\hat{\varphi}^{a}(x_{T})=-\frac{\hat{\rho}^{a}(x_{T})}{\nabla_{T}^{2}}.\label{eq:Poisson}
\end{equation}
The corresponding electric field is then given by
\begin{equation}
E_{i=1,2}^{a}=f(z-t)\partial_{i}\hat{\varphi}^{a}(x_{T}),\qquad E_{3}^{a}=0.\label{eq:ELorenz}
\end{equation}
In Lorenz gauge the chromoelectric fields are purely transverse while
the gauge fields retain only their temporal and longitudinal components.
All fields are nonzero exclusively in the space-time region close
to the light cone, where $J^{a\mu}$ is nonvanishing. In order to
switch to temporal gauge we apply a gauge transformation to the gauge
fields via
\begin{equation}
{A'}_{\mu}^{a}t^{a}=V\left(A_{\mu}^{a}t^{a}+\frac{i}{g}\partial_{\mu}\right)V^{\dagger},\label{eq:transformedfieldnonabelian}
\end{equation}
such that $A_{0}'(x)=0$ is fulfilled at all times. Consequently,
$V$ must satisfy the equation 
\begin{equation}
\frac{\partial}{\partial t}V^{\dagger}=igA_{0}^{a}t^{a}V^{\dagger}.
\end{equation}
Since the gauge field configurations (\ref{eq:Aansatznonabelian})
commute at different times, the solution to this equation does not
require a time-ordered exponential, but is simply given by 
\begin{equation}
V^{\dagger}(t,x,y,z)=\exp\left(ig\hat{\varphi}^{a}(x,y)t^{a}F(t,z)\right)\label{eq:wilson_line}
\end{equation}
with $F(t,z)\equiv\int_{-\infty}^{t}f(z-t')dt'.$ Using this gauge
transformation, the fields in the temporal axial gauge are given by
\begin{equation}
{A'}_{\mu=1,2}^{a}t^{a}=\frac{i}{g}V\left(\partial_{\mu}V^{\dagger}\right),\qquad{A'}_{\mu=0,3}^{a}t^{a}=0.
\end{equation}
The current has to be transformed properly into the temporal axial
gauge ${J'}_{\mu}^{a}t^{a}=V\left(J_{\mu}^{a}t^{a}\right)V^{\dagger}$
as well. The corresponding electric field can be calculated from
$E_{i}^{a}\equiv-\partial^{0}A^{ai}.$ We make two important observations
at this point. In contrast to the situation in Lorenz gauge, the gauge
fields are now purely transversal. Additionally, they are now defined
not only on the nuclear sheet close to the light cone as before, but also
in the spatial region behind each nucleus, forming a trace of constant
gauge fields (see Fig.~\ref{fig:temporal-gauge-overview}). Although
these fields are pure gauge configurations, which can be gauge transformed
to vacuum and thus do not carry any energy, their emergence forces
us to choose fixed boundary conditions in the longitudinal direction. 

The Wilson line (\ref{eq:wilson_line}) required for the transformation
to temporal gauge is completely analogous to the lightlike Wilson
lines used in the boost-invariant formulation \cite{Iancu:2002xk,Iancu:2001yq}.
If we consider $\hat{\rho}^{a}(x_{T})$ as a random variable, the
ansatz (\ref{eq:Jansatznonabelian}) and (\ref{eq:Aansatznonabelian})
leads to uncorrelated fields in the transversal direction, but correlation
over the longitudinal extent of the nucleus. A more general ansatz,
which is beyond the scope of the current work, would also allow for
fluctuations in the longitudinal direction \cite{Fukushima:2007ki}
making a time-ordered exponential in Eq.~(\ref{eq:wilson_line})
necessary.

Furthermore, we introduce infrared (IR) and ultraviolet (UV) regulators
for the solution of the Poisson equation (\ref{eq:Poisson}). This
is done by solving in momentum space:
\begin{equation}
\hat{\varphi}^{a}(k_{T})=\begin{cases}
\frac{\hat{\rho}^{a}(k_{T})}{\left|k_{T}\right|^{2}+m^{2}} & ,\qquad\left|k_{T}\right|\leq\Lambda,\\
0 & ,\qquad\left|k_{T}\right|>\Lambda,
\end{cases}\label{eq:Poisson_reg}
\end{equation}
where the parameters $m$ and $\Lambda$ control the IR and UV regulation
and $\hat{\varphi}^{a}(k_{T})$, $\hat{\rho}^{a}(k_{T})$ are the
Fourier components of $\hat{\varphi}^{a}(x_{T})$ and $\hat{\rho}^{a}(x_{T})$
respectively. The IR regulator $m$ in the expression for $\hat{\varphi}^{a}$
introduces a finite correlation length on the order of $m^{-1}$ in
the transversal directions. The inclusion of the IR regulator and
the UV cutoff does not violate the field equations of motion or the
Gauss constraint, since it can be absorbed into a redefined charge
density 
\begin{equation}
\hat{\rho}'^{a}(k_{T})=\frac{\left|k_{T}\right|^{2}}{\left|k_{T}\right|^{2}+m^{2}}\Theta(\Lambda-\left|k_{T}\right|)\,\hat{\rho}^{a}(k_{T}),
\end{equation}
which satisfies the unmodified Poisson equation in momentum space
\begin{equation}
\hat{\varphi}^{a}(k_{T})=\frac{\hat{\rho}'^{a}(k_{T})}{\left|k_{T}\right|^{2}}.
\end{equation}
Regulating the infrared modes with $m>0$ also enforces global color
neutrality, i.e.
\begin{equation}
\hat{\rho}'^{a}(k_{T}=0)=0.
\end{equation}
On the lattice we initialize the transversal gauge links at $t_{0}-\frac{a_{t}}{2}$
and $t_{0}+\frac{a_{t}}{2}$ via
\begin{equation}
U_{x,i}(t_{0}\pm\frac{a_{t}}{2})=V(t_{0}\pm\frac{a_{t}}{2},x)V^{\t}(t_{0}\pm\frac{a_{t}}{2},x+i),\qquad i\in\{1,2\},\label{eq:initial_conditions_lattice}
\end{equation}
and the longitudinal gauge links are set to the unit element. The
initial electric fields $E_{x,i}(t_{0})$ are computed from the gauge
link update (\ref{eq:eom_4b}). We then evaluate the Gauss constraint
(\ref{eq:gauss_4}) to obtain the correct three-dimensional color
charge density $\rho_{x}(t_{0})$, which is sampled by a number of
particles. One point charge per transverse grid cell is sufficient
to reproduce a given charge density in a transverse plane. The longitudinal
structure requires a higher resolution: In order to obtain a smooth
current with the NGP algorithm, the charge is distributed among $\Np=a_{s}/a_{t}$
particles per cell, which are placed with equal spacing along the
longitudinal direction such that at each time step exactly one particle
crosses a Wigner-Seitz cell boundary. It is not sufficient to divide
the total charge within a cell to the particles equally. The sublattice
distribution of the charges has to be optimized with the charge refinement
algorithm described in the next section.

\subsection{Charge refinement\label{sub:Charge-refinement}}

\begin{figure}
\begin{centering}
\includegraphics[scale=0.65]{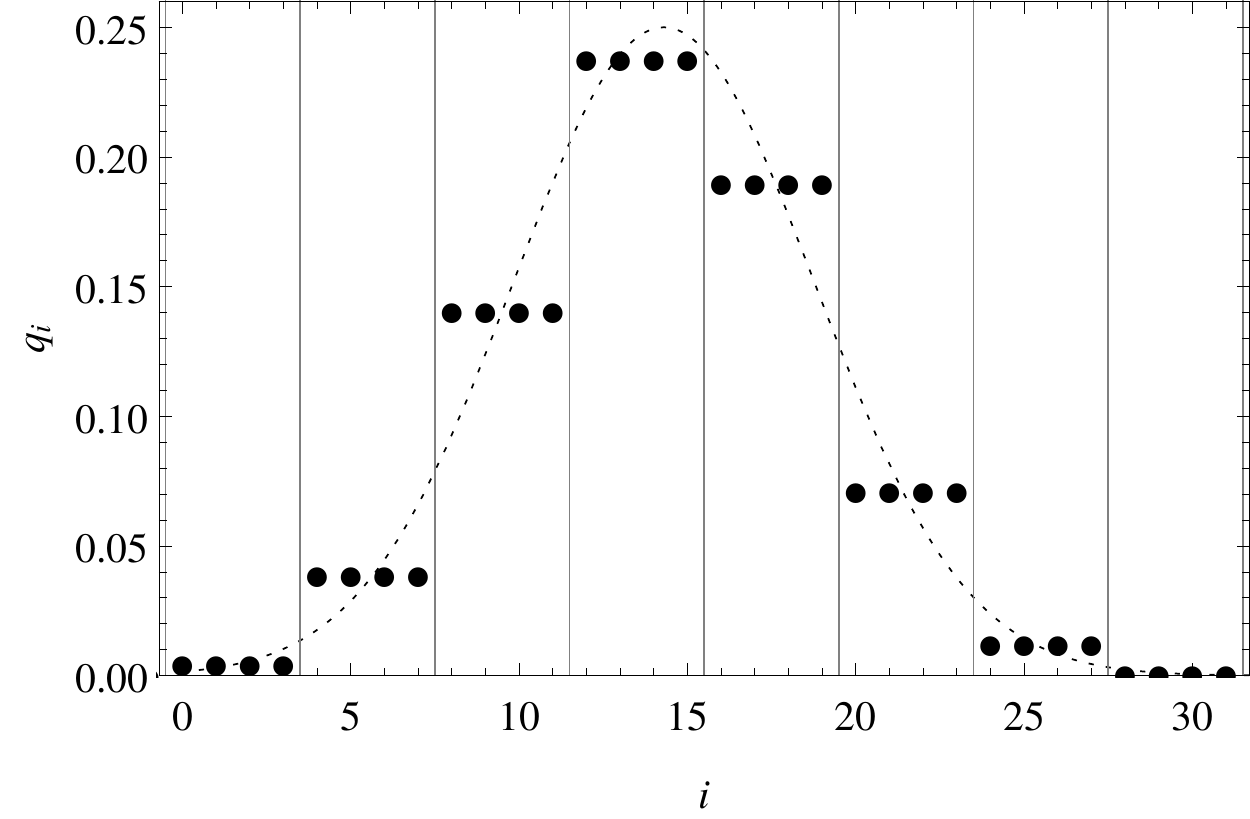}\hfill{}\includegraphics[scale=0.65]{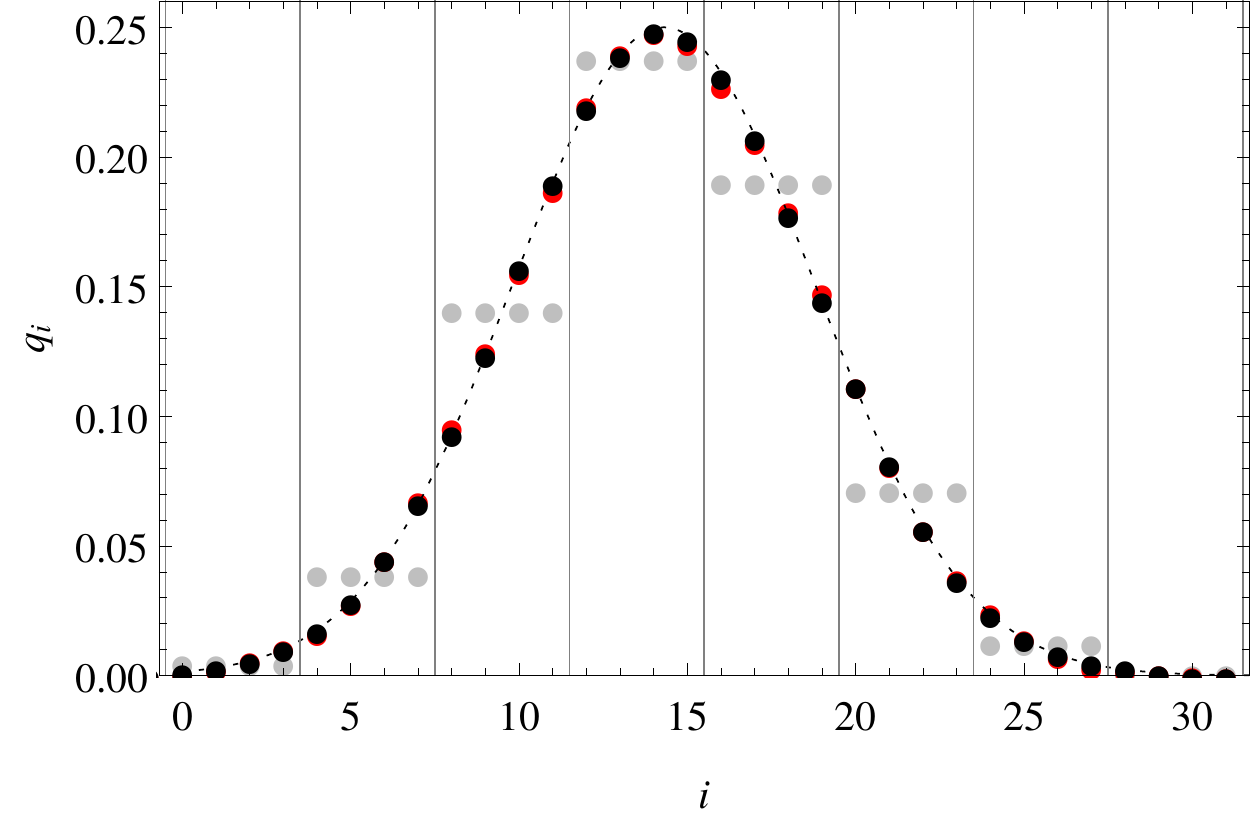}
\par\end{centering}

\protect\caption{Example of charge refinement: the small charge $q_{i}$ is plotted
as a function of the position $i$ for $\protect\Np=4$ small charges
per cell. In the left plot, an initial total charge per Wigner-Seitz
cell (separated by the gray vertical lines and given in this example
by the integral of the dashed line per cell) is equally distributed
among four small charges per cell. After applying the charge refinement
algorithm (right plot), the total charge per Wigner-Seitz cell is
exactly the same as in the left plot, but it approximates the continuous
charge distribution (dashed line) significantly better. Red dots indicate
the result for constant discrete second derivative, Eq.~(\ref{eq:delta_second}),
while black dots show the result for constant discrete fourth derivative
within a cell, Eq.~(\ref{eq:delta_forth}). \label{fig:charge-refinement}}
\end{figure}

Up to this point, we have only specified the total charge in a cell,
but not how the charge is distributed within the cell. A constant
charge distribution within each cell as seen in Fig.~\ref{fig:charge-refinement}
on the left results in a ``jittery'' color current distribution
on the grid over time. This also impacts the evolution of the fields
and in particular leads to spurious longitudinal fields in the direction
of propagation.\footnote{To see why this is the case consider the equation of motion of the
longitudinal chromoelectric field. This argument can already be made
with the Abelian equation $\dot{E}_{L}=(\vec{\nabla}\times\vec{B})_{L}-j_{L}$.
The electric and magnetic fields of a single nucleus moving at the
speed of light are purely transverse, there are no longitudinal components.
Consequently, the longitudinal current must satisfy $j_{L}=(\vec{\nabla}\times\vec{B})_{L}$
at all times at each point in space. Any deviation from this produces
longitudinal chromoelectric fields, which in turn affect the future
time evolution of the other fields. In our simulations the spatial
shape and time behavior of the interpolated current depend on the
sublattice distribution of particle charges. A smooth distribution
of the charges is better at preserving the transversal field structure.} In order to avoid this, the shape of the charge distribution should
be as smooth as possible as seen in Fig.~\ref{fig:charge-refinement}
on the right.

For the NGP algorithm, we have to satisfy that the sum of all small
charges within a Wigner-Seitz cell equals its given total charge $Q_{j}$.
In order to distribute the total cell charge $Q_{j}$ to $\Np$ small
charges $q_{i}$ with $\Np j\leq i<\Np(j+1)$ within the cell, we
can initialize them  according to 
\begin{equation}
q_{i}=\frac{Q_{j}}{\Np},\quad{\rm for\;\;}\Np j\leq i<\Np(j+1).
\end{equation}
We then apply an iterative procedure that ensures that the total charge
within a cell is not altered. At each iteration step, two randomly
chosen neighboring charges $q_{i}$ and $q_{i+1}$ (with $i+1$ not
a multiple of $\Np$) are assigned new values $q'_{i}$ and $q'_{i+1}$
according to
\begin{eqnarray}
q'_{i} & = & q_{i}-\Delta q,\label{eq:q0prime-1}\\
q'_{i+1} & = & q_{i+1}+\Delta q.\label{eq:q1prime-1}
\end{eqnarray}
This ensures for arbitrary $\Delta q$ that the total charge within
the cell is not modified.  If one demands that the discrete second
derivative is constant within a cell, which is equivalent to demanding
that the discrete first derivatives of adjacent points form an arithmetic
series
\begin{equation}
\frac{q'_{i+1}-q'_{i}}{a_{s}}=\frac{1}{2}\left[\frac{q_{i+2}-q_{i+1}}{a_{s}}+\frac{q_{i}-q_{i-1}}{a_{s}}\right],
\end{equation}
then we find
\begin{equation}
\Delta q=\frac{q_{i+2}-3q_{i+1}+3q_{i}-q_{i-1}}{4}.\label{eq:delta_second}
\end{equation}
Applying Eqs.~(\ref{eq:q0prime-1}) and (\ref{eq:q1prime-1}) with
(\ref{eq:delta_second}) repeatedly to all points leads to a convergent
solution that is continuous and piecewise linear in the first derivative.
The algorithm cannot be applied directly to the border of two cells
(i.e.~$i=\Np j-1$), so these points have to be left out.

One can also demand that the discrete third derivatives form an arithmetic
series:
\begin{equation}
\frac{q_{i+2}-3q'_{i+1}+3q'_{i}-q_{i-1}}{a_{s}^{3}}=\frac{1}{2}\Bigg[\frac{q_{i+3}-3q_{i+2}+3q_{i+1}-q_{i}}{a_{s}^{3}}+\frac{q_{i+1}-3q_{i}+3q_{i-1}-q_{i-2}}{a_{s}^{3}}\Bigg].
\end{equation}
On the left-hand side, we use that $q_{i+2}$ and $q_{i-1}$ remain
untransformed. The result is
\begin{equation}
\Delta q=\frac{-q_{i+3}+5q_{i+2}-10q_{i+1}+10q_{i}-5q_{i-1}+q_{i-2}}{12}.\label{eq:delta_forth}
\end{equation}
Convergence is fastest if the results are first iterated according
to condition (\ref{eq:delta_second}) and then according to (\ref{eq:delta_forth}).
An example of this procedure is shown in Fig.~\ref{fig:charge-refinement}.

\subsection{Simulation cycle}

For comprehensiveness we summarize the individual steps in our simulations.
First we generate initial conditions using the methods described in
the last two sections. This includes generating random charge distributions
according to the MV model, solving the two-dimensional Poisson equation
and initializing the chromoelectric fields and gauge links in the
temporal gauge. The Gauss constraint is then used to obtain the charge
density on the grid, which is sampled by a number of colored particles.
The distribution of particle charges is then made smooth with the
charge refinement algorithm. 

After initialization at time $t_{0}$ the known variables are the
particle positions $x(t_{0})$, $x(t_{0}+a_{t})$ and charges $Q(t_{0})$,
the currents $j_{x,i}(t_{0}+\frac{a_{t}}{2})$, the electric fields
$E_{x,i}(t_{0})$ and the gauge links $U_{x,i}(t_{0}+\frac{a_{t}}{2})$.
The variables are then updated as follows:
\begin{enumerate}
\item Compute the new electric field $E_{x,i}(t_{0}+a_{t})$ via Eq.~(\ref{eq:eom_4a})
using $j_{x,i}(t_{0}+\frac{a_{t}}{2})$ and $U_{x,i}(t_{0}+\frac{a_{t}}{2})$.
\item Update $U_{x,i}(t_{0}+\frac{3a_{t}}{2})$ via Eq.~(\ref{eq:eom_4b})
using $E_{x,i}(t_{0}+a_{t})$.
\item Update longitudinal particle positions via Eq.~(\ref{eq:position_update}).
\item Update particle charges $Q(t_{0}+a_{t})$ according to either Eq.~(\ref{eq:right_move_transport})
or Eq.~(\ref{eq:left_move_transport}) (depending on sign of the
particle velocity $v$) if a particle crosses a nearest-grid-point
boundary.
\item Interpolate charge density $\rho_{x}(t_{0}+a_{t})$ using the NGP
scheme and Eq.~(\ref{eq:ngp_charge_interpolation}).
\item Interpolate currents $j_{x,i}(t_{0}+\frac{3a_{t}}{2})$ using either
Eq.~(\ref{eq:right_move_current}) or Eq.~(\ref{eq:left_move_current})
depending on sign of the particle velocity $v$.
\item Compute various observables such as field energy, pressure components,
etc.
\end{enumerate}
This completes a simulation step.

\section{Numerical results\label{sec:Numerical-results}}

\global\long\def\ev#1{\left\langle #1\right\rangle }

For all of our simulations\footnote{The code for our simulation framework is open-source and publicly
available at \url{https://github.com/openpixi/openpixi}.} we use a model similar to the one proposed by McLerran and Venugopalan
\cite{MV1,MV2}. As discussed in Sec.~\ref{sec:Initial-conditions-fields},
we consider charge distributions, which are random in the transversal
direction, but correlated in the longitudinal direction. The randomly
chosen color charge densities $\hat{\rho}_{1,2}^{a}(x_{T})$ in the
transversal plane are taken to be Gaussian with the correlation function
\begin{equation}
\ev{\hat{\rho}_{1,2}^{a}(x_{T})\hat{\rho}_{1,2}^{b}(y_{T})}=g^{2}\mu_{1,2}^{2}\delta^{(2)}(x_{T}-y_{T})\delta^{ab},
\end{equation}
where the parameters $\mu_{1,2}$ control the variance of the fluctuating
charges. McLerran and Venugopalan give an estimate of 
\begin{equation}
\mu^{2}=1.1A^{1/3}\mbox{ fm}^{-2},
\end{equation}
where $A$ is the mass number of the colliding nuclei and the gauge
group is SU(3). For $A=197$ (Au) we get 
\begin{equation}
\mu\approx0.505\mbox{ GeV}.
\end{equation}
We choose $g=2$ as common in CGC literature \cite{Lappi:2003bi,Fukushima:2011nq,Fukushima:2007ki,Lappi:2006hq}.
This leads to a realistic value for the saturation momentum $Q_{s}$,
\begin{equation}
Q_{s}\sim g^{2}\mu\sim2\mbox{ GeV}.
\end{equation}
Even though our simulation is currently restricted to SU(2) for performance
reasons, we still use this value to test our methods in semirealistic
scenarios. 

In all simulations we use Au-Au collisions as the standard case study,
therefore $\mu$ is fixed. However, we vary other parameters such
as the simulation volume, the nucleus width $\sigma$ and IR and UV
regulators. For example, using $N_{T}=128$ cells in the transversal
directions and a lattice spacing of $a=0.028\mbox{ fm}$, the transversal
area roughly covers $12.5\%$ of the area of a single Au nucleus.
In the longitudinal direction we could use $N_{L}=256$ cells, which
covers a length of $5.2\mbox{ fm}$. These are the parameters used
in Sec.~\ref{sub:Comparison-with-boost-invariant}. For other parts
of this paper we chose different parameter sets, which are specified
in the corresponding sections.

We also have to choose the longitudinal thickness $l$ of the nuclei,
which is controlled by the longitudinal Gaussian width $\sigma$.
In Sec.~\ref{sub:Comparison-with-boost-invariant} we show that $\sigma=4a_{s}$
is a good lower limit to avoid lattice artifacts. We approximate the
thickness of the Gaussian profile by
\begin{equation}
l\approx4\sigma.
\end{equation}
Comparing the thickness $l$ to the longitudinal extent of the Lorentz contracted
nucleus $\frac{2R_{A}}{\gamma}$ , we obtain an estimate for the gamma
factor $\gamma$.
\begin{equation}
\gamma=\frac{2R_{A}}{4\sigma}=\frac{R_{A}}{8a_{s}},
\end{equation}
where $R_{A}\approx1.25A^{1/3}\mbox{ fm}$ is the radius of a nucleus.
With the values from above we get
\begin{equation}
\gamma\approx45.
\end{equation}
This value for $\gamma$ corresponds to a center-of-mass energy of
$\sqrt{s_{NN}}\approx90\,{\rm GeV}$, however in the course of our
paper we will demonstrate results obtained for $\gamma=11-455$, corresponding
to an energy range of $\sqrt{s_{NN}}=20-850\mbox{ GeV}$. This energy
range contains in particular parts of the parameter space explored
by the low-energy Au+Au collisions at RHIC in the beam energy scan
program with center-of-mass energies between $\sqrt{s_{NN}}=7.7-62.4\mbox{ GeV}$
\cite{Adamczyk:2013gw}.

For the solution of the Poisson equation in the transversal plane
we employ IR and UV regularization as in Eq.~(\ref{eq:Poisson_reg}).
Infrared regularization leads to average color neutrality and suppresses
long-range forces (e.g.~monopoles and dipoles) on length scales $m^{-1}$.
This is used to include effects of confinement in the classical simulation.
The confinement radius is roughly $1\mbox{ fm}$, therefore one possibility
is to set
\begin{equation}
m\approx(1\mbox{ fm})^{-1}\approx200\mbox{ MeV}\approx\Lambda_{\mbox{QCD}}.
\end{equation}
However we also work with values of up to $2$ GeV to study the dependence
of observables on the IR regulation. The UV cutoff $\Lambda$ is introduced
to eliminate high-momentum modes in the transversal plane whose dispersion
relation on the lattice differs from the analytic case. Unless otherwise
noted we use $\Lambda=10$ GeV.

\subsection{Comparison with boost-invariant initial conditions\label{sub:Comparison-with-boost-invariant}}

\begin{figure}
\begin{centering}
\subfloat{\protect\centering{}\protect\includegraphics{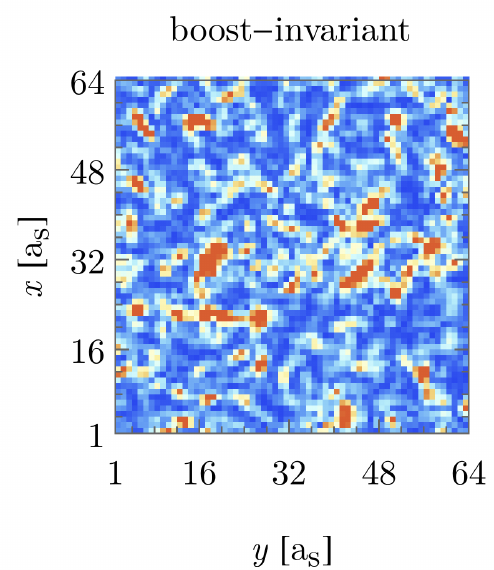}\protect}\hfill{}\hfill{}\subfloat{\protect\centering{}\protect\includegraphics{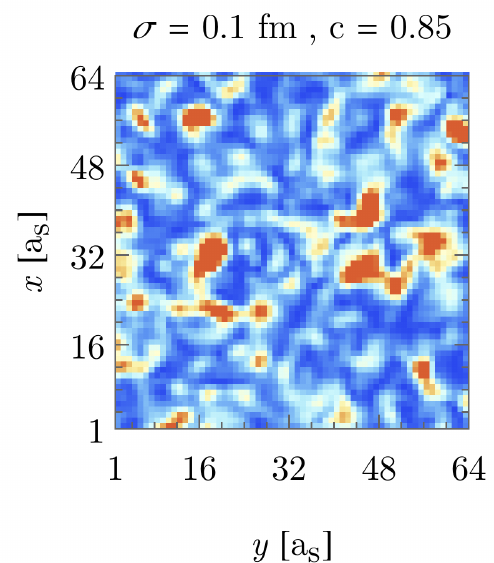}\protect}\hfill{}\hfill{}\subfloat{\protect\centering{}\protect\includegraphics{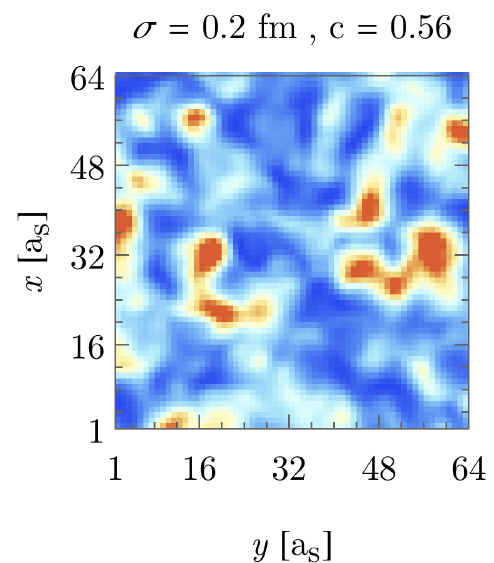}\protect}
\par\end{centering}

\protect\caption{Density plot of the energy density component $\mbox{tr}E_{L}^{2}(x_{T})$
as a function of the transverse coordinate $x_{T}=(x,y)$ in the center
region of the collision for a single event. The left panel shows the
boost-invariant (``analytic'') result for $\tau=0^{+}$. The middle
and right panels show the simulation results for two different values
of the thickness parameter $\sigma$. The correlation coefficient
$c$ quantifies how similar the energy density distributions are to
the boost-invariant case. Thinner nuclei (middle) lead to a correlation
coefficient of $c=0.85$, whereas the energy density distributions
of thicker nuclei (right) are more washed out with lower values of
$c=0.56$. The grid size $N_{L}\cdot N_{T}^{2}$ of the simulation
is set to $256\cdot128^{2}$ with a lattice spacing of $a_{s}=0.028\,\mbox{fm}$
and a time step of $a_{t}=a_{s}/2$. The IR regulator is set to $m=2\,\mbox{GeV}$
and the UV cutoff is $\Lambda=10\,\mbox{GeV}$. The transversal area
covers $12.5\%$ of the full area of a single Au nucleus, but we only
show a quarter of the area to make the similarities visually more
obvious. \label{fig:Density-plot-correlation}}
\end{figure}

It is important to check if the results produced by our 3+1 dimensional
simulations are similar to 2+1 dimensional boost-invariant simulations,
at least in the limit of thin nuclei. In the boost-invariant formulation
it is natural to work with proper time $\tau=\sqrt{t^{2}-z^{2}}$
and rapidity $\eta=\frac{1}{2}\ln\frac{t+z}{t-z}$ as coordinates
for the forward light cone, where the collision event at $t=z=0$
is used as the origin of the coordinate system. Note however that
in collisions of nuclei with finite thickness, there is some ambiguity
involved in choosing the space-time coordinates ($t_{c},z_{c}$) of
the collision. As a definition we set $(t_{c},z_{c})$ to the space-time
point of the maximum overlap of the Gaussian longitudinal profiles.
The main advantage of this definition is that these coordinates are
independent of the thickness parameter $\sigma$. To verify the agreement
with the boost-invariant case, we compare the longitudinal chromoelectric
fields created in our simulations to the fields, which are used as
initial conditions for boost-invariant simulations. The boost-invariant
initial conditions for the electric field at $\tau=0^{+}$ created
by the collision of charge densities of two nuclei $\hat{\rho}_{1}(x_{T})$
and $\hat{\rho}_{2}(x_{T})$ are given by \cite{Krasnitz:1998ns}
\begin{equation}
\left.E_{L}(x_{T})\right|_{\tau=0^{+}}=-ig\sum_{i=1,2}\left[\alpha_{1}^{i}(x_{T}),\alpha_{2}^{i}(x_{T})\right],\label{eq:boost_invariant_EL}
\end{equation}
where $\alpha_{1,2}^{i}(x_{T})$ is determined from the relations
\begin{eqnarray}
e^{iga_{s}\alpha_{1,2}^{i}(x_{T})} & = & e^{ig\hat{\varphi}_{1,2}^{a}(x_{T})t^{a}}e^{-ig\hat{\varphi}_{1,2}^{a}(x_{T}+i)t^{a}},\label{eq:bi_ic_1}\\
\Delta_{T}\hat{\varphi}_{1,2}^{a}(x_{T}) & = & -\hat{\rho}_{1,2}^{a}(x_{T}),\label{eq:bi_ic_2}
\end{eqnarray}
which are similar to our initial conditions in the laboratory frame:
The first equation corresponds to Eq. (\ref{eq:initial_conditions_lattice}),
and  the second one to the Poisson equation (\ref{eq:Poisson_eq}).
This result is obtained in the Fock-Schwinger gauge $\tau A^{\tau}=0$. 

\begin{figure}
\begin{centering}
\subfloat{\protect\centering{}\protect\includegraphics{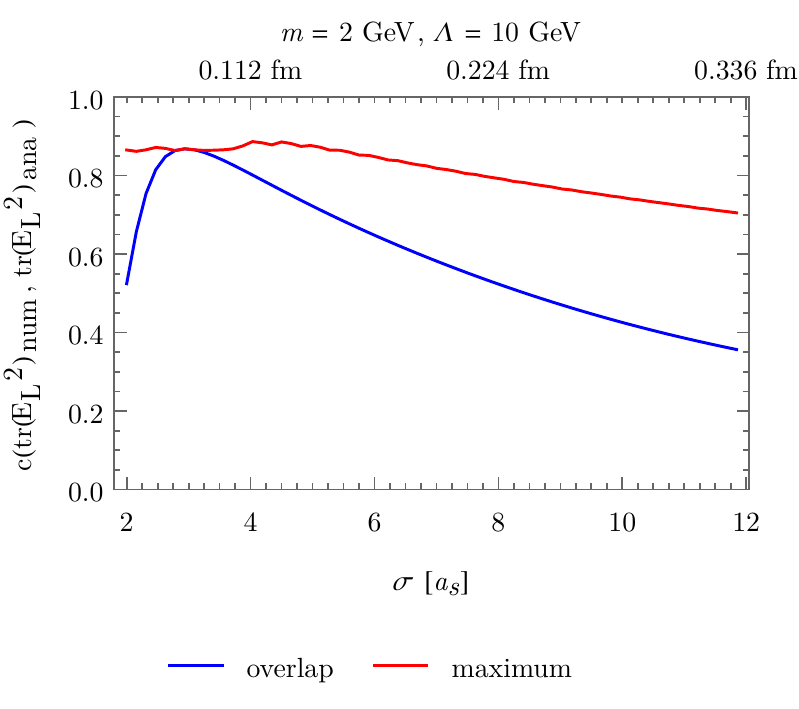}\protect}\hspace{0.5cm}\subfloat{\protect\centering{}\protect\includegraphics{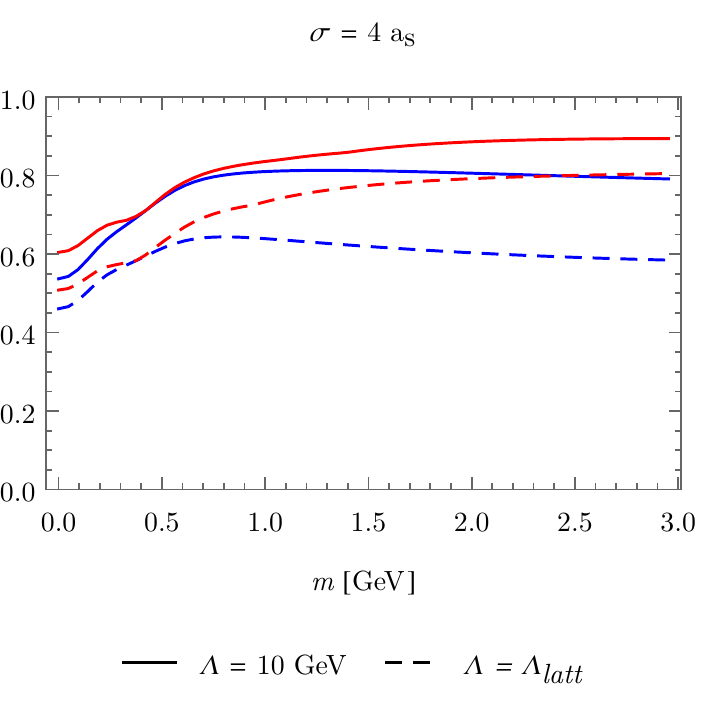}\protect}
\par\end{centering}

\protect\caption{Comparison of simulations to boost-invariant initial conditions. This
plot shows the correlation coefficient of $\mbox{tr}E_{L}^{2}(x_{T})_{\mathrm{num}}$
in the central region with the boost-invariant result for $\mbox{tr}E_{L}^{2}(x_{T})_{\mathrm{ana}}$
at $\tau=0^{+}$ as a function of $\sigma$ and $m$. A correlation
coefficient of $1$ implies perfect agreement between the numerical
and the analytical result. The blue solid line shows the correlation
when the nuclei completely overlap and the red line is the maximum
correlation achieved during the evolution. The correlation increases
for thinner nuclei. Small values of $m$ and $\sigma$ lead to decreased
correlations due to numerical instabilities, which appear at high
field amplitudes. The simulation parameters are the same as in Fig.$\;$\ref{fig:Density-plot-correlation}
except that we vary the thickness parameter $\sigma$ and IR regulator
$m$. \textit{Left panel:} Correlations as a function of the thickness
parameter $\sigma$.\label{fig:Correlations-width} \textit{Right
panel:} Correlations as a function of the IR regulator $m$. For the
thick curves we used $\Lambda=10\,\mbox{GeV}$ as a UV cutoff. Dashed
lines use the UV cutoff $\Lambda_{latt}$ given by the lattice. \label{fig:Correlations-infrared}\label{fig:Comparison-of-simulations}}
\end{figure}

In our case the simulations start before the collision, where the
nuclei are well separated in the longitudinal direction such that
the gauge field in the center between them vanishes to numerical accuracy.
The longitudinal chromoelectric fields which we want to compare to
Eq.~(\ref{eq:boost_invariant_EL}) are produced by numerically evolving
the fields of the nuclei. We test our simulation as follows: We generate
two initial charge densities $\hat{\rho}_{(1,2)}(x_{T})$ and compute
$\left.\mbox{tr}E_{L}^{2}(x_{T})\right|_{\tau=0^{+}}$, which is a
gauge-invariant expression. Then we use the same charge densities
to run a 3+1 dimensional simulation with some finite nuclear thickness
controlled by the Gaussian width $\sigma$. We record the energy density
contribution of the longitudinal electric field $\mbox{tr}E_{L}^{2}(x_{T})$
as a function of the transversal coordinate $x_{T}$ during the collision
in the central region $\eta=0$. We then compute the correlation coefficient
$c$ between the numerical (simulation) result $\mbox{tr}E_{L}^{2}(x_{T}){}_{\mathrm{num}}$
and the analytic (boost-invariant) expression $\mbox{tr}E_{L}^{2}(x_{T}){}_{\mathrm{ana}}$
via
\begin{equation}
c(\mbox{tr}(E_{L}^{2})_{\mathrm{num}},\mbox{tr}(E_{L}^{2}){}_{\mathrm{ana}})\equiv\frac{\mbox{cov}(\mbox{tr}(E_{L}^{2})_{\mathrm{num}},\mbox{tr}(E_{L}^{2})_{\mathrm{ana}})}{\sigma_{\mathrm{num}}\sigma_{\mathrm{ana}}},
\end{equation}
where the covariance across the transversal plane is defined by 
\begin{equation}
\mbox{cov}(\mbox{tr}(E_{L}^{2})_{\mathrm{num}},\mbox{tr}(E_{L}^{2})_{\mathrm{ana}})=\sum_{x_{T}}\left(\mbox{tr}E_{L}^{2}(x_{T}){}_{\mathrm{num}}-\overline{\mbox{tr}(E_{L}^{2})}_{\mathrm{num}}\right)\left(\mbox{tr}E_{L}^{2}(x_{T}){}_{\mathrm{ana}}-\overline{\mbox{tr}(E_{L}^{2})}_{\mathrm{ana}}\right),
\end{equation}
with the mean values $\overline{\mbox{tr}(E_{L}^{2})}_{(\mathrm{num},\mathrm{ana})}$
and the standard deviations $\sigma_{(\mathrm{num},\mathrm{ana})}$
associated with $\mbox{tr}E_{L}^{2}(x_{T}){}_{\mathrm{num}}$ and
$\mbox{tr}E_{L}^{2}(x_{T}){}_{\mathrm{ana}}$ respectively. The mean
and standard deviation are understood to be computed across the transversal
plane. A plot of the energy densities $\mbox{tr}E_{L}^{2}(x_{T})$
for different widths is shown in Fig.~\ref{fig:Density-plot-correlation}.

The correlation between the numerical and analytical results is recorded
as a function of time, the nuclear thickness $\sigma$ and the UV
cutoff $\Lambda$. The results for a single event as a function of
$\sigma$ are shown in Fig.~\ref{fig:Correlations-width} (left panel):
The blue (lower) curve corresponds to the fixed time $t_{c}$ where
the two nuclei overlap completely. We see that the correlation increases
for thinner widths $\sigma$, but at a certain point around $\sigma\approx3a_{s}$
the correlation is reduced due to discretization errors. Very thin
longitudinal profiles tend to disperse, produce unphysical longitudinal
fields even before the collision and eventually become unstable. This
is because thin widths cannot be properly resolved on the lattice
below a certain threshold. To ensure numerical stability we deduce
a minimum width of $\sigma_{min}=4a_{s}$ for the nuclear thickness
in our simulations. We remark that very fine lattices with small (in
physical units) lattice spacings are required to accurately simulate
thin nuclei on the lattice.

The red (upper) curve in Fig.$\;$\ref{fig:Comparison-of-simulations}
on the left is the maximum value of the correlation during the collision.
We see that thicker nuclei also produce fields which are similar to
the boost-invariant case (thus leading to higher correlations), but
at earlier times than $t_{c}$. This happens because they start to
overlap much earlier, producing the characteristic longitudinal electric
fields. The time evolution from the onset of the overlap to the full
overlap at $t_{c}$ changes the fields, resulting in low values of
the correlations at $t_{c}$ . 

In Fig.~\ref{fig:Correlations-infrared} (right panel) we study the
effects of the IR regulator $m$ and the UV cutoff $\Lambda$ on our
results by fixing the width $\sigma$ and varying the values of $m$
and $\Lambda$. We see that cutting off high momentum modes whose
dispersion relation differs from the continuum case increases the
correlation with the analytic result. Regulating the UV modes becomes
necessary because in the MV model all available modes in momentum
space are populated up to the lattice cutoff scale $\Lambda_{\mathrm{latt}}$.
Increasing the resolution of the simulation box without regulating
the UV modes does not lead to any improvement.

We also observe that the correlation coefficient is largely independent
of the IR regulator $m$. However, lower values of $m$ decrease the
correlation significantly in the same manner as small values of $\sigma$
do. Small $m$ boosts the amplitudes of the low momentum modes of
$\hat{\varphi}^{a}(x_{T})$ [as is apparent from Eq.~(\ref{eq:Poisson_reg})],
which drives the same numerical instability we see when using very
small values of $\sigma$. This instability can be cured by using
finer grids (i.e.~smaller lattice spacings $a_{s}$) while keeping
the volume of the simulation box and all other physical parameters
fixed. A smaller lattice spacing $a_{s}$ for the same physical volume
of the box brings the gauge links $U_{x,i}$ closer to the group identity
element $\mathbf{1}$ and consequently the lattice approximations
of the fields become more accurate. We note that this instability
is of numerical nature only and also appears in the evolution of a
single nucleus without any collision.

Studying the correlation between our numerical results and the analytic
expressions for the boost-invariant initial conditions shows that
we are able to correctly describe boost-invariant collisions in the
limit of thin nuclei. However it also reveals that one has to be careful
in choosing simulation parameters, in particular $\sigma\gtrsim4a_{s}$.
To describe Au-Au collisions in our simulation framework we work with
an IR regulator of $m=2\,\mbox{GeV}$ (which is of the order of the
saturation momentum) and a UV cutoff $\Lambda=10\,\mbox{GeV}$ (which
is used to cut off high momentum modes not satisfactorily described
on the lattice). These parameters are used in the following sections
unless otherwise noted.

\subsection{Pressure anisotropy\label{sub:Pressure-anisotropy}}

\begin{figure}
\subfloat{\protect\centering{}\protect\includegraphics{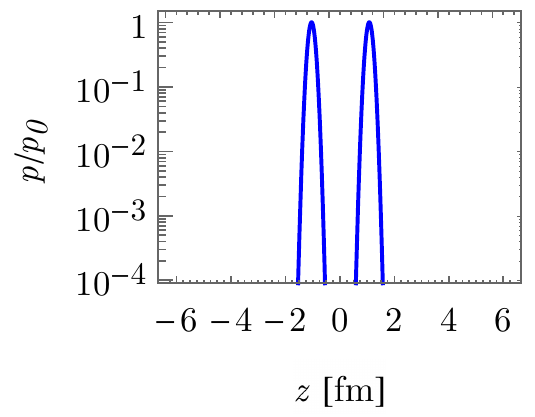}\protect}\hfill{}\subfloat{\protect\centering{}\protect\includegraphics{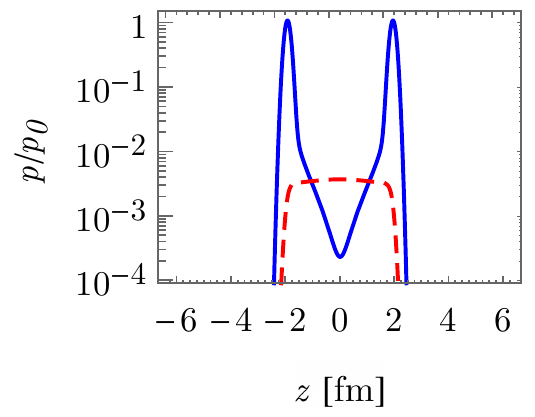}\protect}\hfill{}\subfloat{\protect\centering{}\protect\includegraphics{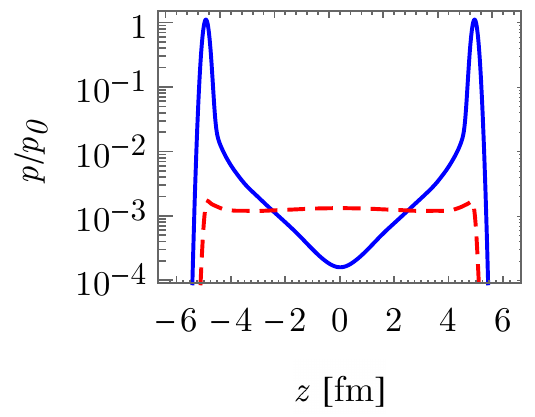}\protect}

\protect\caption{Longitudinal and transverse pressure components as functions of the
longitudinal coordinate $z$ in the laboratory frame at different
times $t$ before and after the collision. The coordinate origin is
centered around the collision event at $t=0$ and $z=0$. The blue
curve describes the longitudinal pressure $p_{L}(z)$ and the red
dashed curve is the transverse pressure component $p_{T}(z)$. The
longitudinal chromoelectric and chromomagnetic fields characteristic
for the glasma contribute to the transverse pressure $p_{T}$. The
pressure components are normalized to the maximum longitudinal pressure
$p_{0}$ of the initial nuclei. For these plots we use a grid size
of $320\times256^{2}$ with a lattice spacing of $a_{s}=0.04\,\mbox{fm}$
and a time step of $a_{t}=\frac{a_{s}}{2}$. The thickness parameter
is set to $\sigma=4a_{s}$ (which corresponds to a gamma factor of
$\gamma\approx23$), the IR regulator is set to $m=2\,\mbox{GeV}$
and the UV cutoff is set to $\Lambda=10\,\mbox{GeV}$. \textit{Left
panel:} Before the collision: $t=-1\,\mbox{fm}/c$. \textit{Middle
panel:} After the collision: $t=2\,\mbox{fm}/c$. \textit{Right panel:}
$t=5\,\mbox{fm}/c$. \label{fig:pL-pT-labframe}}
\end{figure}

\global\long\def\e{\varepsilon}
A prominent phenomenon in the early stages of heavy-ion collisions
is the pressure anisotropy of the glasma fields and the subsequent
isotropization of the system. The main observables in this context
are the transversal and longitudinal pressure components $p_{T}=\e_{L}$
and $p_{L}=\e_{T}-\e_{L}$ with the longitudinal and transversal energy
density components given by
\begin{eqnarray}
\e_{L} & = & \frac{1}{2}\left(E_{z}^{a}E_{z}^{a}+B_{z}^{a}B_{z}^{a}\right),\\
\e_{T} & = & \frac{1}{2}\sum_{i=x,y}\left(E_{i}^{a}E_{i}^{a}+B_{i}^{a}B_{i}^{a}\right).
\end{eqnarray}
Our simulation framework enables us to compute the pressure components
as functions of time $t$ and the longitudinal and transverse coordinates
$z$ and $x_{T}$. To simplify we average over $x_{T}$, which is
natural within the MV model. A plot of the pressure components in
the laboratory frame at different times is shown in Fig.~\ref{fig:pL-pT-labframe}.
The initially purely transverse fields of the incoming nuclei manifest
themselves as large Gaussian bumps in the longitudinal pressure component.
During the collision the transverse pressure component builds up and
remains largely flat afterwards. The longitudinal pressure in the
laboratory frame falls off exponentially towards the center of the
collision. In order to better compare our results to the boost-invariant
case it is sensible to switch to the comoving frame described by proper
time $\tau=\sqrt{t^{2}-z^{2}}$ and rapidity $\eta=\frac{1}{2}\ln\frac{t+z}{t-z}$.
We choose the space-time coordinates $(t_{c},z_{c}$) of the collision
as in Sec.$\;$\ref{sub:Comparison-with-boost-invariant}.

\begin{figure}
\begin{centering}
\includegraphics[scale=0.9]{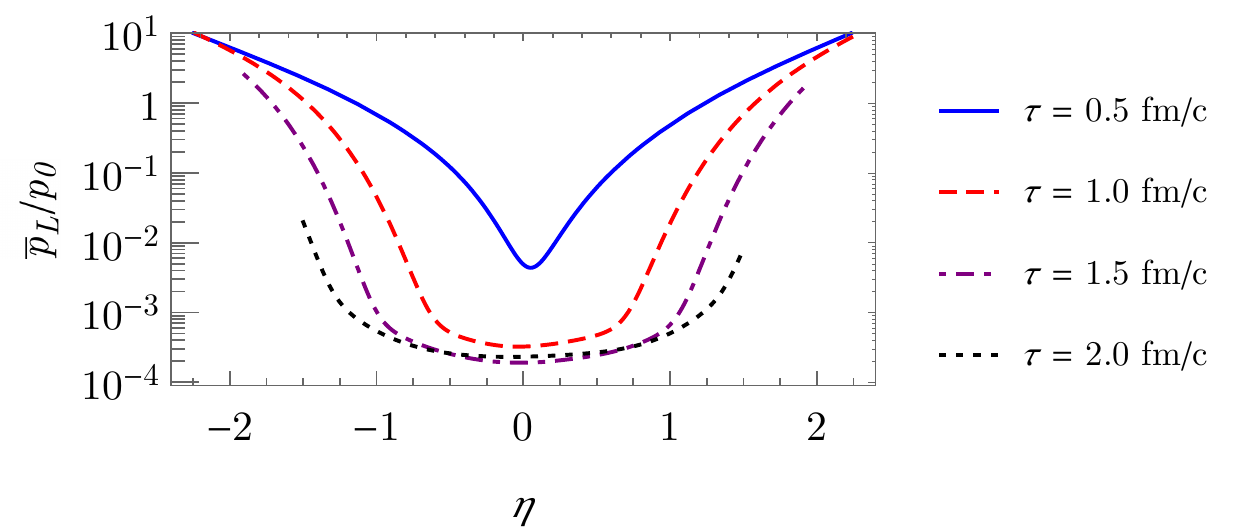}
\par\end{centering}

\protect\caption{Longitudinal pressure component $\bar{p}_{L}(\eta)$ in the comoving
frame as a function of rapidity $\eta$ for different proper times
$\tau$. At later times the longitudinal pressure becomes flat within
the rapidity interval $(-1,1)$. Even though the nuclei in this simulation
are relatively thick ($\gamma\approx23$) we still recover approximate
boost invariance. For these plots we use the same simulation parameters
as in Fig.~\ref{fig:pL-pT-labframe}. \label{fig:pL_comoving}}
\end{figure}

By introducing the longitudinal component of the Poynting vector
\begin{equation}
S_{L}\equiv2\mathrm{tr}\left(\vec{E}\times\vec{B}\right)_{z},
\end{equation}
we can compute the transformed longitudinal pressure
\begin{equation}
\bar{p}_{L}(\tau,\eta)=p_{L}(\tau,\eta)\cosh^{2}\eta+\varepsilon(\tau,\eta)\sinh^{2}\eta-2S_{L}(\tau,\eta)\cosh\eta\sinh\eta.
\end{equation}
The transverse pressure component is unaltered by the coordinate transformation.
A plot of the longitudinal pressure $\bar{p}_{L}(\tau,\eta)$ in the
comoving frame is shown in Fig.~\ref{fig:pL_comoving}. It reveals
that at early times $\bar{p}_{L}$ is still largely influenced by
the tails of the colliding nuclei. At later times $\bar{p}_{L}$ becomes
flat in the midrapidity region, which is consistent with approximate
boost invariance. We have to keep in mind that within our simulations
the observables we compute are always slightly influenced by the initial
fields of the nuclei, especially at early proper times.

We now turn towards studying the pressure anisotropy. For the further
analysis it will be sufficient to stay in the central region $\eta=0$.
In the boost-invariant case the initial glasma fields at $\tau=0^{+}$
are made of purely longitudinal color flux tubes, which leads to highly
anisotropic initial pressures $\left.p_{T}\right|_{\tau=0^{+}}=\left.\e_{L}\right|_{\tau=0^{+}}$
and $\left.p_{L}\right|_{\tau=0^{+}}=-\left.\e_{L}\right|_{\tau=0^{+}}$.
As the flux tubes expand, they generate transversal electric and magnetic
fields until $\e_{L}\simeq\e_{T}$ and $p_{L}\simeq0$ \cite{Fujii:2008dd}.
This is the free-streaming limit observed in boost-invariant CGC simulations
and stands in contrast to the observation of an isotropized quark-gluon
plasma where $p_{T}\simeq p_{L}$ after a few $\mbox{fm}/c$ \cite{Romatschke:2007mq,Ryblewski:2012rr}.
It has been shown that boost invariance breaking fluctuations drive
instabilities in the glasma, which can move the system towards isotropization
\cite{Gelis:2013rba,Fukushima:2011nq,Berges:2012cj}. In our simulations
we explicitly violate boost invariance by introducing a finite nucleus
thickness. It is therefore interesting to investigate the effects
of the thickness parameter $\sigma$ on the pressure anisotropy of
the glasma.

\begin{figure}
\begin{centering}
\subfloat{\protect\centering{}\protect\includegraphics{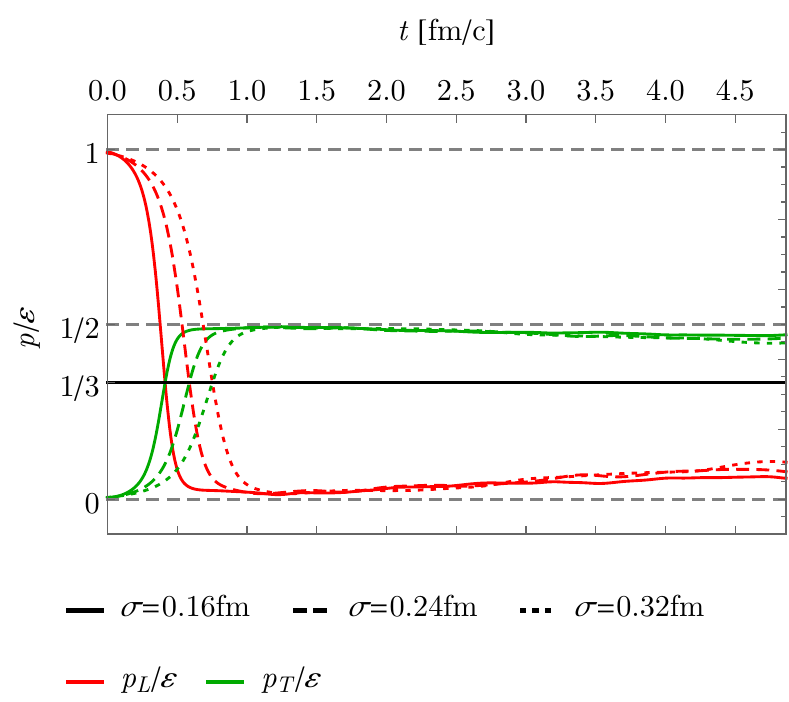}\protect}\hfill{}\subfloat{\protect\centering{}\protect\includegraphics{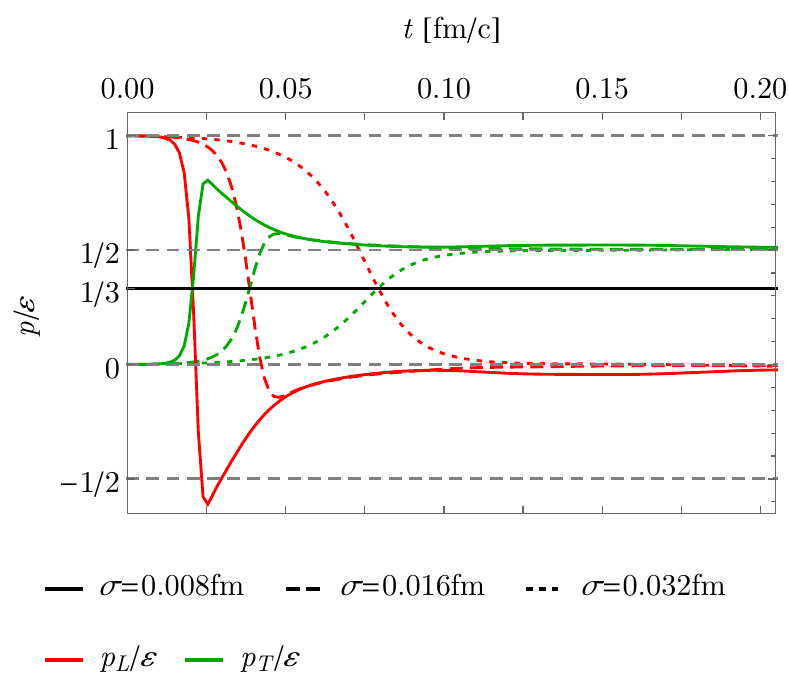}\protect}
\par\end{centering}

\protect\caption{Longitudinal and transversal pressure components in the central region
$\eta=0$ as a function of time for various nuclear thicknesses $\sigma$.
An IR regulator of $m=2\,\mbox{GeV}$ and a UV cutoff of $\Lambda=10\,\mbox{GeV}$
has been used. The detailed simulation parameters are explained in
the Sec.$\;$\ref{sub:Pressure-anisotropy}. \textit{Left panel:}
Pressure components for thick nuclei.\label{fig:Pressure-components-thick}
\textit{Right panel:} Pressure components for thin nuclei.\label{fig:Pressure-components-thin}\label{fig:pressure-comparison}}
\end{figure}

For our numerical studies it is convenient to introduce the pressure
to energy density ratios $\frac{p_{T}}{\e}$ and $\frac{p_{L}}{\e}$
with $\e=\e_{L}+\e_{T}$. The free-streaming limit then corresponds
to $\frac{p_{T}}{\e}\simeq\frac{1}{2}$ and $\frac{p_{L}}{\e}\simeq0$.
Isotropization would be signaled by $\frac{p_{T}}{\e}\simeq\frac{p_{L}}{\e}\simeq\frac{1}{3}$.
Both the pressure and energy density components are averaged over
the transverse plane and $32$ events are used for the statistical
sampling. We choose a grid size of $320$ cells in the longitudinal
direction and $256^{2}$ cells to resolve the transversal area. For
collisions of thick nuclei in Fig.~\ref{fig:Pressure-components-thick}
(left panel) we choose a lattice spacing of $a_{s}=0.04\mbox{ fm}$.
The transversal grid then covers the full area $\pi R_{A}^{2}$ of
a gold nucleus. 

For simulations of thin nuclei in Fig.~\ref{fig:Pressure-components-thin}
(right panel) we are forced to use smaller lattice spacings of $a_{s}=0.008\,\mbox{fm}$
(for $\sigma=0.032\,\mbox{fm}$), $a_{s}=0.004\,\mbox{fm}$ (for $\sigma=0.016\,\mbox{fm}$)
and $a_{s}=0.002\,\mbox{fm}$ (for $\sigma=0.008\,\mbox{fm}$), because
grids much larger than $320\times256^{2}$ as used here currently
exceed our available computational resources. The transversal area
then only covers $4\%$, $1\%$ and $0.25\%$ of the full area respectively.
The temporal spacing is set to $a_{t}=\frac{a_{s}}{2}$.

The results are shown in Fig.~\ref{fig:pressure-comparison} and
there are several observations we make:
\begin{enumerate}
\item From Fig.~\ref{fig:Pressure-components-thick} (left panel) we see
that we recover the free-streaming limit of the boost-invariant case.
Isotropization is not reached within possible simulation times due
to limitations from both the longitudinal and transversal simulation
box size. We can observe slight movement of both pressure components
towards the desired value of $\frac{1}{3}$, but not within any realistic
time scales.
\item The initial pressures directly after the collision behave differently
compared to the boost-invariant case. In our simulations of thick
nuclei in Fig.~\ref{fig:Pressure-components-thick} (left panel)
we see that in the beginning $p_{L}$ dominates $p_{T}$ due to the
presence of the transverse fields of the colliding nuclei. As the
nuclei recede from the collision volume the created glasma fields
have already reached the free-streaming limit and therefore no negative
longitudinal pressures are observed. 
\item In the results for thin nuclei in Fig.~\ref{fig:Pressure-components-thin}
(right panel) we can recover negative longitudinal pressure. The colliding
nuclei move away from the collision center fast enough, leaving behind
longitudinal color flux tubes, which have not decayed yet. The still
largely longitudinal fields generate negative pressure, which is characteristic
for the early glasma phase. 
\end{enumerate}
We remark here that our ansatz for the initial conditions relies
on an assumption about the longitudinal structure of the nuclei. The
initial conditions described in Sec.~\ref{sec:Initial-conditions-fields}
imply correlation of the charge density in the longitudinal direction
of order $\sigma$ and correlation in the transversal direction of
order $m^{-1}$. The charge distribution is random in the transversal
direction, but there is no random longitudinal structure. As a consequence
we were able to drop the time ordering in Eq.~(\ref{eq:wilson_line}).
However, it has been shown that random longitudinal structure (which
demands proper path/time ordering) in the initial nucleus fields -
among other effects - leads to higher initial energy densities in
the glasma \cite{Fukushima:2007ki}. Additional longitudinal randomness
might also give rise to larger deviations from the boost-invariant
case after the collision. This could be similar to boost invariance
breaking perturbations of the glasma, which cause plasma instabilities
that have been found to accelerate isotropization \cite{Gelis:2013rba,Fukushima:2011nq,Berges:2012cj}.
It is therefore conceivable to expect that implementing this random
longitudinal structure in our initial conditions will change the results
and could lead to faster isotropization times. Detailed understanding
of these issues requires an analysis of gluon occupation numbers in
momentum space and their temporal behavior. We plan to investigate
this in a future publication.

\subsection{Energy production\label{sub:Energy-production}}

\begin{figure}[t]
\begin{centering}
\subfloat{\protect\centering{}\protect\includegraphics[scale=0.35]{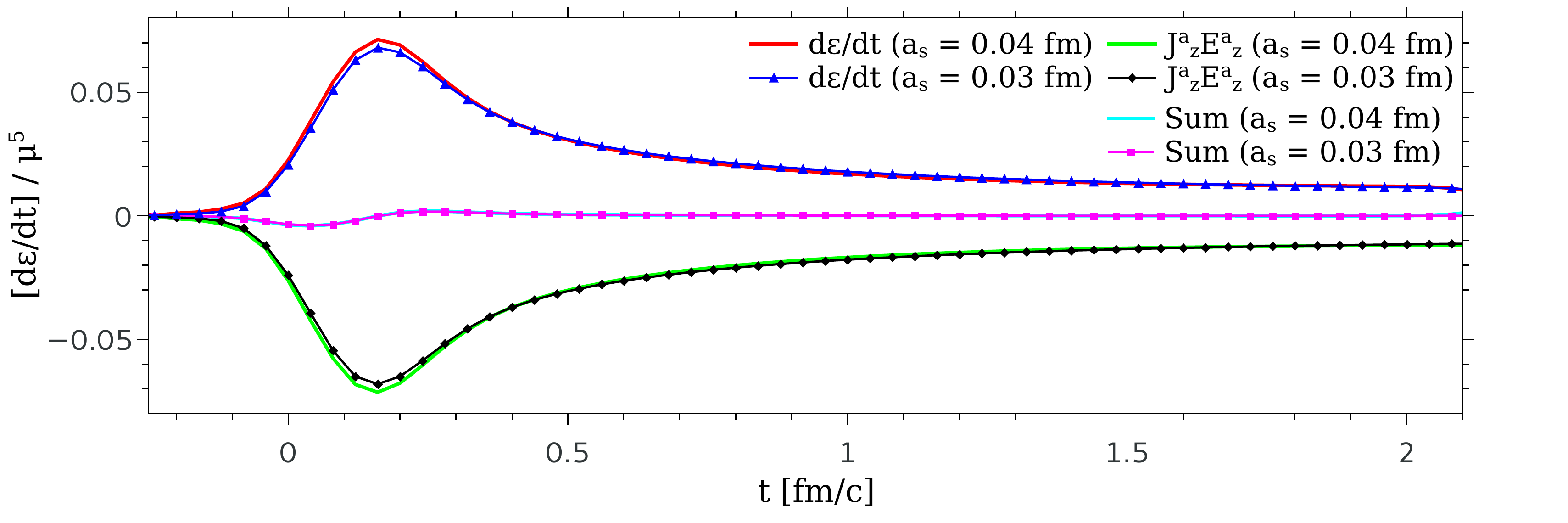}\protect}\hfill{}\subfloat{\protect\centering{}\protect\includegraphics[scale=0.35]{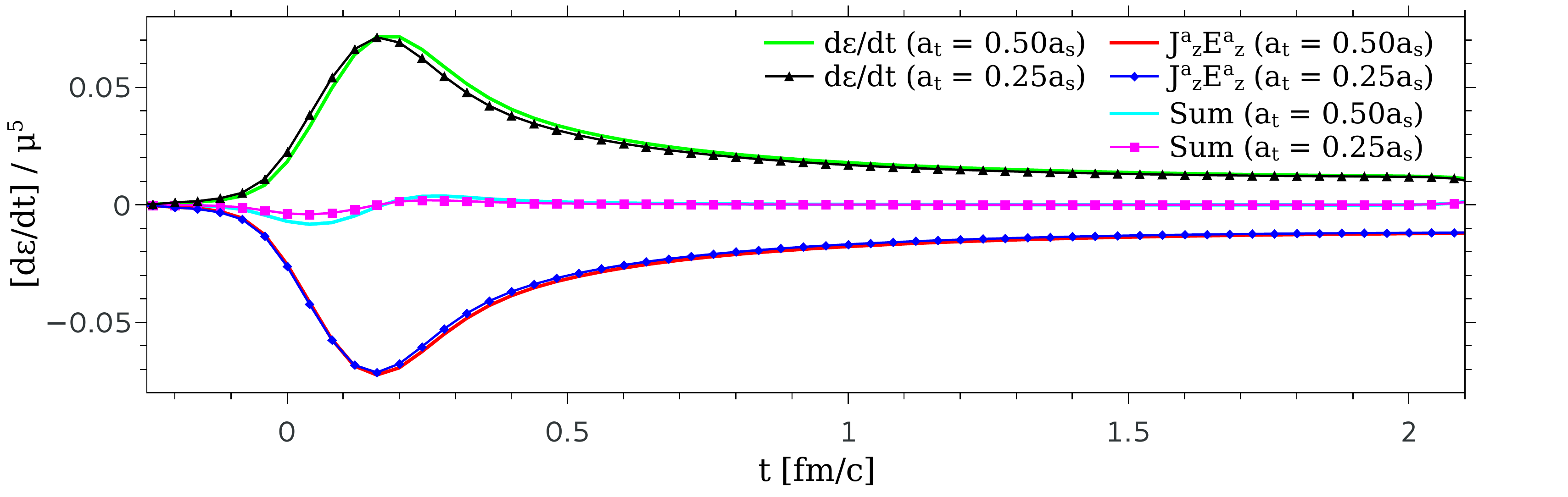}\protect}
\par\end{centering}

\protect\caption{Energy production as a function of time with different spatial and
temporal discretizations. The ``Sum'' curves correspond to the left-hand
side of Eq.~(\ref{eq:continuity1}). Their deviation from zero is
a consequence of lattice artifacts and can be reduced by using finer
time discretizations. The results have been obtained on a cubic lattice
with a fixed volume of $(5.12\:\mathrm{fm})^{3}$ with the IR regulator
set to $m=1\,\mbox{GeV}$ and an UV cutoff $\Lambda=10\,\mbox{GeV}$.
The nuclear thickness $\sigma$ was set to $0.16\,\mbox{fm}$. We
averaged over ten configurations in order to have a sufficient statistical
sample. \textit{Upper panel:} Varying spatial discretization by keeping
$a_{t}=0.01\:\mathrm{fm}/c$ fixed.\label{fig:Energy-production-1}
\textit{Lower panel:} Varying temporal discretization by keeping $a_{s}=0.04\:\mathrm{fm}$
fixed.\label{fig:Energy-production-2}\label{fig:Energy-production}}
\end{figure}

One of the fundamental assumptions made in the CGC framework is the
separation of hard and soft degrees of freedom, which are modeled
as external color charges and classical gauge fields respectively.
As a result of the collision there is an energy exchange between the
charges and the fields. However, since the nuclei are assumed to be
recoilless, the hard sector acts as an inexhaustible energy reservoir
for the gauge fields. The resulting field energy increase can be interpreted
as the work done by the charges against the field. In the boost-invariant
case this effect is implicitly included in the initial conditions
for the fields at $\tau=0^{+}$. In our approach we are able to explicitly
compute the energy increase during and after the collision. The change
of the total field energy density $\e$ as a function of time can
be formulated in terms of an energy continuity equation
\begin{equation}
\frac{d\varepsilon}{dt}+\frac{1}{V}\int\partial_{i}S_{i}d^{3}x+\frac{1}{V}\int E_{i}^{a}J_{i}^{a}d^{3}x=0,\label{eq:continuity1}
\end{equation}
which is the non-Abelian version of the Poynting theorem. The time
dependence of $\e$ is governed by two terms: the components $S_{i}$
of the Poynting vector $S_{j}\equiv2\mathrm{tr}\left(\vec{E}\times\vec{B}\right)_{j}$
and $E_{i}^{a}J_{i}^{a}$. The integral over the total derivative
of the Poynting vector can be omitted in the continuum. On the lattice
this term only gives a negligible contribution due to discretization
errors. In the scalar product $E_{i}^{a}J_{i}^{a}$ the only nonvanishing
part of the current $J_{i}^{a}$ is the longitudinal component $J_{z}^{a}$
and therefore the expression reduces to $E_{z}^{a}J_{z}^{a}$. Consequently,
the energy production is caused by longitudinal chromoelectric fields
in the glasma and must be centered around the collision event and
the boundary of the forward light cone where the color currents are
nonzero. 

The energy increase as a function of time is shown in Fig.~\ref{fig:Energy-production-1}.
We observe that the total energy density is conserved before the onset
of the collision when the external charges and the classical fields
describing both nuclei are propagating through vacuum. Afterwards
there is a strong energy increase during as well as after the collision.
At later times there is an ongoing, but slowly decreasing energy production,
which finally becomes almost constant. To check the stability of our
results with respect to a change in the spatial and temporal resolution
of the grid we vary the spatial lattice spacing $a_{s}$ and the time step
$a_{t}$. Overall there is a good agreement between results at different
discretizations. The violation of Eq. (\ref{eq:continuity1}) is small
and can be further reduced using smaller time steps as seen in the
lower plot of Fig.$\;$\ref{fig:Energy-production}.

\subsection{Suppression of longitudinal chromomagnetic fields}

\begin{figure}
\begin{centering}
\subfloat{\protect\centering{}\protect\includegraphics{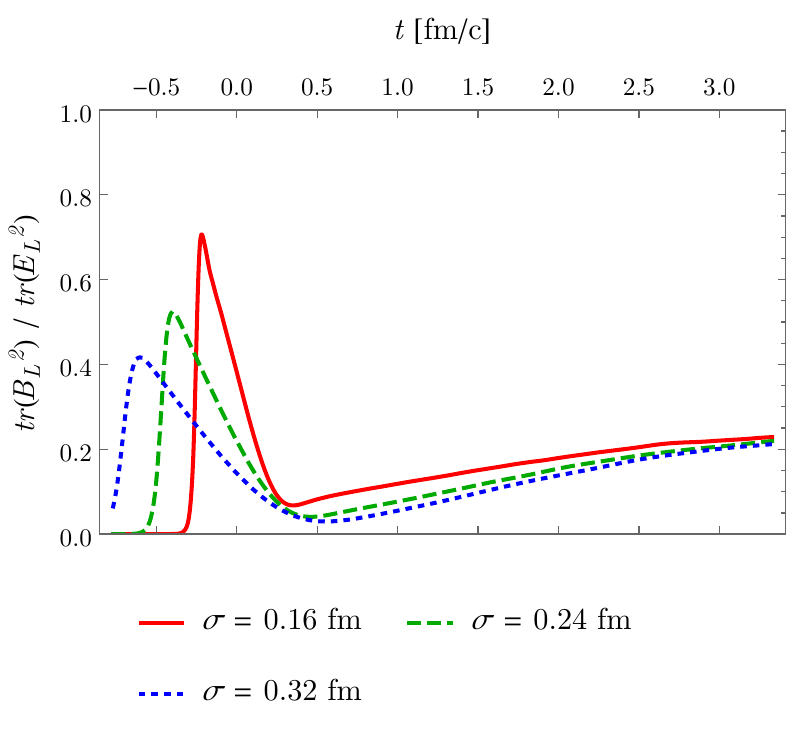}\protect}\hfill{}\subfloat{\protect\centering{}\protect\includegraphics{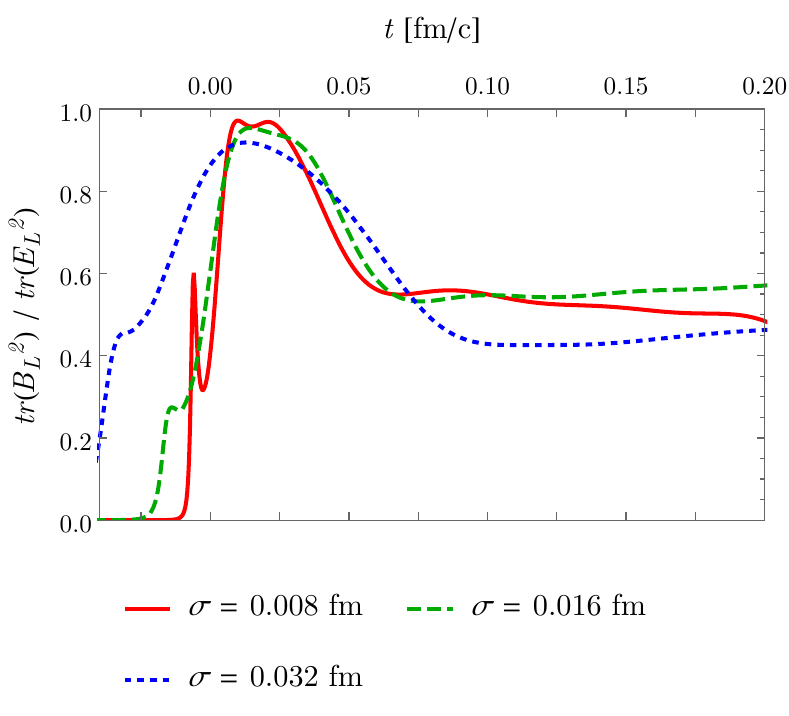}\protect}
\par\end{centering}

\protect\caption{Ratio of magnetic to electric longitudinal energy density contributions
as a function of time for various nuclear thicknesses $\sigma$. The
ratio increases for thin nuclei, but magnetic flux tubes are still
heavily suppressed compared to the boost-invariant scenario. The simulation
parameters are the same as in Sec.$\;$\ref{sub:Pressure-anisotropy}.
\textit{Left panel:} Ratio of longitudinal energy density components
for thick nuclei. \label{fig:Ratio-thick} \textit{Right panel:} Ratio
of longitudinal energy density components for thin nuclei. \label{fig:Ratio-thin}\label{fig:ratio-energy-density}}
\end{figure}

\begin{figure}
\begin{centering}
\includegraphics{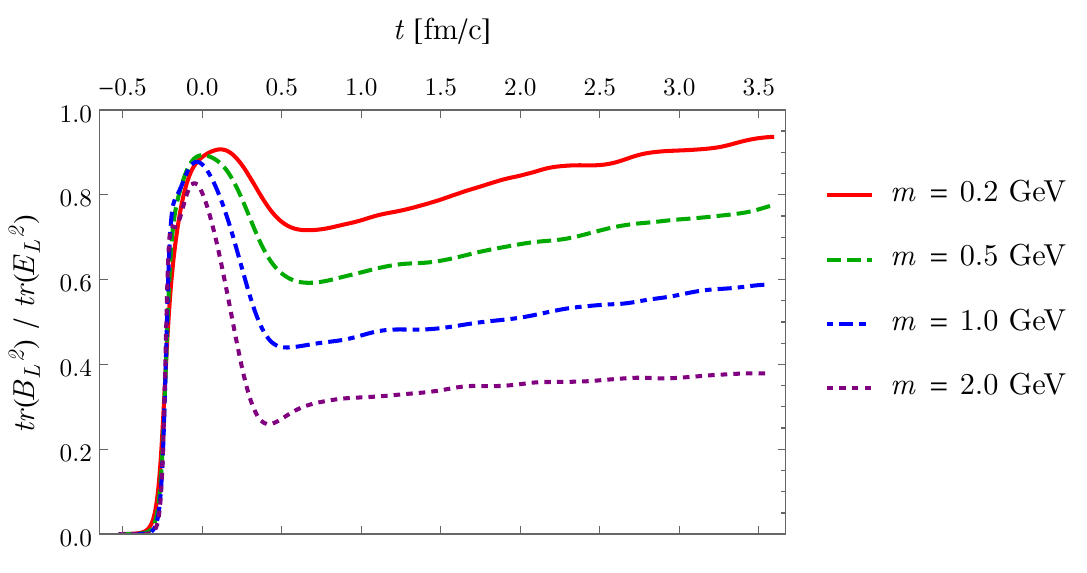}
\par\end{centering}

\protect\caption{Ratio of magnetic to electric longitudinal energy density contributions
as a function of time for various values of the IR regulator $m$.
For this plot we used a grid size of $256^{3}$ cells with a lattice
spacing $a_{s}=0.02\,\mbox{fm}$ and averaged over $32$ events. The
transversal area covers $25\%$ of the full area of a gold nucleus.
The thickness parameter is set to $\sigma=0.08\,\mbox{fm}$, which
corresponds to $\gamma\approx45$. We approach the boost-invariant
limit for small $m$. \label{fig:ratio-energy-density-IR}}
\end{figure}

In the following we investigate the production of longitudinal chromomagnetic
fields $B_{L}^{a}$ and chromoelectric fields $E_{L}^{a}$ characteristic
for the glasma at early times. In the boost-invariant case the contributions
to the energy density from magnetic and electric color flux tubes
($\mbox{tr}B_{L}^{2}$ and $\mbox{tr}E_{L}^{2}$ respectively) should
be equal after averaging over initial conditions. In our simulations
with finite $\sigma$ we observe that this is not the case and there
is a dependency on the thickness parameter $\sigma$ as well as the
IR regulator $m$. The results are presented in Figs.~\ref{fig:ratio-energy-density}
and \ref{fig:ratio-energy-density-IR}. 

Figure \ref{fig:Ratio-thick} shows the ratio of magnetic and electric
longitudinal fields $\ev{\mbox{tr}B_{L}^{2}}/\ev{\mbox{tr}E_{L}^{2}}$
in the central region ($\eta=0$) for a range of values of the nucleus
thickness $\sigma$ and an IR regulator $m=2\,\mbox{GeV}$. Collisions
of thick nuclei show a very small ratio of about $0.1-0.2$ after
the collision. In the case of thin nuclei in Fig.~\ref{fig:Ratio-thin}
(right panel) the ratio increases to roughly $\sim0.5$, which is
still far away from the ``canonical'' value of $1$ in the boost-invariant
scenario. Note that due to the small physical volumes used in the
simulations of thin nuclei it is harder to achieve adequate statistics.
As a result, the curves in Fig.~\ref{fig:Ratio-thin} in the right
panel are not as smooth as in the left panel.

The results do not only depend on the thickness $\sigma$. In Fig.~\ref{fig:ratio-energy-density-IR}
the results are shown for a fixed nuclear thickness $\sigma=0.08\,\mbox{fm}$
($\gamma\approx45$) and a varying IR regulator\footnote{We remark that varying the IR regulator $m$ has only a weak influence
on the pressure anisotropy. }. We observe that reducing $m$ to $200\,\mbox{MeV}$ (which roughly
corresponds to a correlation length of the color fields of the order
of the confinement radius $1\,\mbox{fm}$) leads to better agreement
with the boost-invariant case with a ratio of $\sim0.8$. Note that
this dependency of the ratio of magnetic and electric longitudinal
fields on the IR regulator $m$ is not present in the boost-invariant
initial conditions \cite{Krasnitz:1998ns}.

The presented results seemingly suggest a suppression of chromomagnetic
flux tubes (or an overproduction of chromoelectric flux tubes) in
the glasma phase when introducing a finite nucleus thickness. However,
the strong dependency of the magnetic to electric longitudinal field
ratio on the IR regulator $m$ leads us to suspect that this discrepancy
between our simulations and the boost-invariant case is an artifact,
which can be attributed to the initial conditions introduced in Sec.~\ref{sec:Initial-conditions-fields}.
As already mentioned there and in Sec.~\ref{sub:Pressure-anisotropy}
the longitudinal structure of our nuclei does not include ``longitudinal
randomness''. Consequently, the typical color structures in our initial
conditions have a thickness proportional to $\sigma$ and a transversal
width of the order of $m^{-1}$. To be consistent with the picture
of a highly Lorentz contracted nucleus modeled by classical Yang-Mills
fields one would demand that $\sigma m\ll1$, such that nucleons within
the nucleus are also contracted to flat ``pancakes''. Therefore,
if we move away from the limit $\sigma m\ll1$, we can expect to see
deviations from the boost-invariant case, but these deviations may
very well solely be due to the longitudinal coherence. This reasoning
is consistent with our simulation results: In the case of a thickness
of $\sigma=0.16\,\mbox{fm}$ with an IR regulator of $m=2\,\mbox{GeV}$
the longitudinal magnetic fields are weakened as seen in Fig.~\ref{fig:Ratio-thick}.
Here we have a value of $\sigma m=1.6$, which corresponds to color
structures which are prolonged in the longitudinal direction. We can
compare this to the case of $\sigma=0.08\,\mbox{fm}$ and $m=200\,\mbox{MeV}$
as presented in Fig.~\ref{fig:ratio-energy-density-IR}. The ratio
of magnetic to electric fields is closer to $1$ and at the same time
we have $\sigma m=0.08$, which can be considered small. 

Including random longitudinal structure in the nuclei as suggested
in \cite{Fukushima:2007ki} will help to clarify, if suppressed 
longitudinal magnetic fields are a physical consequence of a finite
thickness or if the suppression is just an artifact of our ansatz.
However with the reasoning presented above we suspect that this effect
will disappear for more realistic initial conditions.

\section{Conclusion\label{sec:Conclusion}}

In this work we have simulated heavy-ion collisions in the laboratory
frame with thick nuclei in the McLerran-Venugopalan model. Finite
thickness in the longitudinal direction allows the simulation of collisions
at lower energies, but requires abandoning boost invariance in the
calculation as well as including nontrivial color source evolution
in the simulation. Both can be readily implemented using CPIC in the
laboratory frame. With our framework we are able to access a range
of nuclear thicknesses down to those corresponding to center-of-mass
energies as used in the low-energy beam energy scan program of RHIC
and up to LHC energies. 

We started from an analytic solution of a non-Abelian random color
current sheet of finite extent in the longitudinal direction and the
corresponding field configuration that propagate at the speed of light.
The discretization of this solution on a grid requires refining the
charge distribution on sublattice resolution. For the interpolation
between particles and fields, we utilized the nearest-grid-point method
as a charge conserving interpolation scheme. A distinct feature of
our approach is the possibility to explicitly compute the energy,
which is pumped into the Yang-Mills fields by the propagating color
charges. We verified that the energy increase correctly satisfies
the Poynting theorem for non-Abelian fields.

We compared calculations in the laboratory frame with results from
boost-invariant approaches. Concentrating on gauge-invariant observables,
we see that the correlation of initial conditions right after the
collision increases for thinner nuclei, which means that boost invariance
is restored in this limit. We computed the components of the energy-momentum
tensor, especially focusing on the pressure parallel and perpendicular
to the propagation direction. We show that our pressure distributions
in laboratory frame coordinate space correspond to largely rapidity independent
pressure distributions in Bjorken coordinates for the midrapidity
region. Our results confirm the previous findings, which established
the picture of strongly pronounced pressure anisotropy during the
very early phase of the fireball evolution. 

For thicker nuclei we find the following deviations: There is a suppression
of the chromomagnetic longitudinal components of the energy-momentum
tensor with respect to their chromoelectric counterparts. We analyzed
this phenomenon and determined its dependence on the thickness parameter
and the IR regulator. Regarding pressure components, we observe a
slow tendency towards isotropization in our simulations. A more detailed
future investigation including random longitudinal structure in our
model could potentially further reduce isotropization times.

 Other possible and planned improvements are the extraction of particle
spectra in order to compare with experimentally measured multiplicities
and also some rather technical aspects, like improved interpolation
prescriptions, which could be beneficial to widen the scope of parameters
accessible to our numerical approach. Another step towards a more
realistic simulation of QCD processes in heavy-ion collisions at low
collision energies would be the inclusion of backreaction of the classical
gauge fields onto the color charges. In the future the CPIC framework
could also allow us to take interactions and scatterings between the
hard constituents of both nuclei into account. Such steps, however,
would go beyond the usual assumptions of the CGC effective theory
and may require an improved understanding of the internal nuclear
structure.

\section{Acknowledgments}

This work has been supported by the Austrian Science Fund FWF, Project No. P 26582-N27 and Doctoral program No. W1252-N27. The computational results presented have
been achieved using the Vienna Scientific Cluster (VSC). We would
like to thank Anton Rebhan, S\"{o}ren Schlichting, Andreas Schmitt
and Raju Venugopalan for useful discussions.

\bibliographystyle{utphys}
\phantomsection\addcontentsline{toc}{section}{\refname}\bibliography{labframe-arxiv-only,labframe-published_nourl}

\end{document}